\documentclass[final,authoryear,3p,times,twocolumn,fleqn]{elsarticle} 

\usepackage{amssymb}
\usepackage{amsmath}
\usepackage{textcomp}

\journal{Earth and Planetary Science Letters}

\begin{document}

\begin{frontmatter}


\title{Melting and Mixing States of the Earth's Mantle after the Moon-Forming Impact}


\author{Miki Nakajima}
\ead{mnakajima@caltech.edu}
\author{David J. Stevenson}
\address{Division of Geological and Planetary Sciences,
California Institute of Technology, 1200 E California Blvd., MC 150-21, Pasadena, CA 91125, USA.}

\begin{abstract}
The Earth's Moon is thought to have formed by an impact between the Earth and an impactor around 4.5 billion years ago.
This impact could have been so energetic that it could have mixed and homogenized the Earth's mantle. However, this view appears to be inconsistent with geochemical studies that suggest that the Earth's mantle was not mixed by the impact.
Another plausible outcome is that this energetic impact melted the whole mantle, but the extent of mantle melting is not well understood even though it must have had a significant effect on the subsequent evolution of the Earth's interior and atmosphere.
To understand the initial state of the Earth's mantle, we perform giant impact simulations using smoothed particle hydrodynamics (SPH) for three different models: (a) standard: a Mars-sized impactor hits the proto-Earth, (b) fast-spinning Earth: a small impactor hits a rapidly rotating proto-Earth, and (c) sub-Earths: two half Earth-sized planets collide.
We use two types of equations of state (MgSiO$_3$ liquid and forsterite) to describe the Earth's mantle.
We find that the mantle remains unmixed in (a), but it may be mixed in (b) and (c). 
The extent of mixing is most extensive in (c).
Therefore, (a) is most consistent and (c) may be least consistent with the preservation of the mantle heterogeneity, while (b) may fall between. 
We determine that the Earth's mantle becomes mostly molten by the impact in all of the models. 
The choice of the equations of state does not affect these outcomes.
Additionally, our results indicate that entropy gains of the mantle materials by a giant impact cannot be predicted well by the Rankine-Hugoniot equations.
Moreover, we show that the mantle can remain unmixed on a Moon-forming timescale if it does not become mixed by the impact.

\end{abstract}
\end{frontmatter}

%


\section{Introduction}
\label{intro}
The so-called giant impact hypothesis is a widely accepted explanation for the origin of the Moon \citep{HartmannDavis1975, CameronWard1976}.  
According to the standard version of this hypothesis, a Mars-sized impactor hit the proto-Earth and created a disk around the planet from which the Moon accreted.
This hypothesis has been favored because it can explain the Moon's mass, iron depletion, and the angular momentum of the Earth-Moon system.
However, this model has difficulty in explaining the fact that the Earth and Moon have nearly identical isotopic ratios (e.g. oxygen, silicon, and tungsten, \citealt{Wiecheretal2001, Herwartzetal2014, Armytageetal2012, Toubouletal2007}). The typical outcome of a giant impact simulation is that the disk materials are derived mainly from the impactor \citep[e.g.,][]{Benzetal1986}, which is often assumed to have had different isotopic ratios given that the oxygen isotopic ratios between the Earth and Mars differ by $0.321 \text{\textperthousand}$ \citep{Franchietal1999}.
\cite{Reuferetal2012} report that an impact at a high impact velocity and steep impact angle would deliver more Earth's mantle materials to the disk, but it is still difficult to explain the identical isotopic ratios.

\cite{PahlevanStevenson2007} have suggested that turbulent mixing in the Earth's atmosphere and the disk homogenized the isotopic ratios of the two reservoirs. 
This model could potentially solve the isotopic problem, but it has several shortcomings. 
This mechanism may not work for all of the observed isotope systems, such as Si \citep{Pahlevanetal2011, Armytageetal2012}. 
Furthermore, this mixing would have required the Earth's whole mantle convection during the Moon formation, but such convection may not have occurred due to the thermally stratified structure of the mantle after the impact (discussed in Sections \ref{model:mix} and \ref{mixing}). 
Even if the post-impact mixing caused the disk to have the same isotopic reservoir as, say, the outer 80\% of the mantle but failed to equilibrate with the inner 20\%, then the Earth and Moon could still be isotopically different if there had been a subsequent mixing of the Earth's mantle after the Moon formation. 
At present, there is no detectable oxygen isotopic difference (with respect to the three isotopes, which are referred to as $\Delta^{17} {\rm O} \equiv \delta^{17}{\rm O}-0.52\delta^{18}{\rm O}$) among Earth rocks. 

The possibility that the impactor had a similar oxygen isotopic ratio to the Earth has been recently revisited.
\cite{KaibCowan2015} investigate feeding zones of terrestrial planets based on orbital calculations and predict that this possibility is $\sim 5 \%$ or less.
\cite{MastrobuonoBattistietal2015} have performed similar analyses and report that the possibility could be as high as $20-40\%$ if its standard deviation ($\pm \sigma$) is included, although the most plausible value of this possibility ($\sim 10-20 \%$) may not be significantly different from the value estimated by \cite{KaibCowan2015}.  
It is also possible that the impactor could have been compositionally similar to enstatite chondrites \citep[e.g.,][]{Herwartzetal2014}, whose compositions are much more similar to those of the Earth than those of Mars. 
Alternatively, a recent model of planet formation, the so-called Grand Tack model \citep{Walshetal2011}, may suggest a different outcome. 
This Grand Tack model suggests that the planetesimal disk was truncated at 1 AU due to migration of gas giants and, as a result, terrestrial planets mainly formed from the inner part of the disk.
This model may increase the chance of having an impactor with a composition similar to the Earth (personal communications with Alessandro Morbidelli).
This increased probability may arise possibly because the main source of the Earth's materials was confined to a limited region of distances from the Sun, or perhaps because of more efficient mixing of the source materials than the standard models predict (discussed in Section \ref{radmix}). In either case, this finding would imply that the Earth is different from Mars but not necessarily different from the terminal giant impacting body that led to the formation of the Moon (often called as ``Theia'').
It should be noted that having the same oxygen isotopic ratios for the proto-Earth and impactor does not necessarily explain the nearly identical tungsten and silicon isotopic ratios of the two \citep{Dauphasetal2014, FitoussiBourdon2012}.

New giant impact models have been suggested as alternatives.
\cite{CukStewart2012} propose that an impactor hit a rapidly rotating proto-Earth (called the ``fast-spinning Earth"), whereas \cite{Canup2012} suggests a giant impact between two half-Earth-mass objects (here we call this ``sub-Earths", and the model is also called ``large impactor'' and ``half Earth'' in other literatures). In these cases, the composition of the disk would have been similar to that of the Earth's mantle; therefore, the models could explain the isotopic similarities. 
In these models, the angular momentum of the Earth-Moon system after the impact was 2-3 times as large as today's value.
\cite{CukStewart2012} suggest that the evection resonance between the Moon and the Sun could have transferred the excess angular momentum of the Earth-Moon system to the Sun-Earth system.  
This resonance occurs when the period of precession of the pericenter of the Moon is equal to the Earth's orbital period \citep{ToumaWisdom1998}. 
It is also possible that there is some other resonance that yields the same result \citep{WisdomTian2015}. 
Nonetheless, it is not yet clear whether any resonance could efficiently remove the excess of the angular momentum (personal communications with Jack Wisdom). 
The existence of a resonance does not imply the removal of a large amount of angular momentum because that would depend on a particular and possibly narrow choice of tidal parameters.

These new models are indistinguishable in terms of the oxygen isotopic ratios, but additional geochemical constraints may differentiate these models. 
For example, it has been suggested that the Earth's mantle may not have been completely mixed by the giant impact. 
This has been drawn from various isotopic studies, especially those on the Hf-W system.
Hf is a lithophile (``rock-loving") element, whereas W is a moderately siderophile (``iron-loving") element. 
$^{182}$Hf decays to $^{182}$W with a 9 Myr half-life; thus, the mantle of a planet would have an enhanced $^{182}$W/$^{184}$W if differentiation occurred while Hf was still alive (within the first $\sim60$ Myr after CAI formation).
Most terrestrial rocks have similar values of $^{182}$W/$^{184}$W \citep{LeeHalliday1996}, 
but \cite{Willboldetal2011} and \cite{Toubouletal2012} find that 2.8 and 3.8 billion years old rocks have ratios in excess of these values.  
This finding may indicate that the early mantle was heterogeneous (while $^{182}$Hf  was still present) and that the signature was preserved at least until 2.8 billion years ago. 
Determining the cause of the heterogeneity is an active area of research. 
It should be noted that the chemical heterogeneity could take many forms, including a non-uniformity of noble gas mole fraction \citep[e.g.,][]{Mukhopadhyay2012} or a discrete layer of denser mantle material at the base of the mantle because of the formation of a basal magma ocean \citep[e.g.,][]{Labrosseetal2007}.

Regardless of the cause or form, if the mantle heterogeneity formation predated the Moon-forming impact, the Earth's mantle may not have been mixed even by the giant impact. Previous studies based on the lunar $^{182}$W/$^{184}$W ratio suggest that the Moon formed as early as 30 Myr after CAI formation \citep[e.g.,][]{Leeetal1997}, whereas \cite{Toubouletal2007} propose that $\sim$60\% crystallization of the lunar magma ocean occurred after $\sim$60 Myr by taking into account the excess of $^{182}$W formed by neutron capture of $^{181}$Ta.
This is consistent with the age estimate based on other systems (e.g., Sm-Nd, \citealt{CarlsonLugmair1988}, and recent studies on orbital dynamics, \citealt{Jacobsonetal2014}). 
Although the age of the Moon is still under debate, recent work tends to suggest a young age of the Moon ($> 60$ Myr). Herein, we focus on the scenario in which the mantle heterogeneity predated the formation of the Moon.

In addition to the mantle mixing, understanding the extent of the impact-induced mantle melting is important because the extent affected the evolution of the Earth's interior and atmosphere \citep[e.g.,][]{AbeMatsui1986, TonksMelosh1993, ElkinsTanton2008}. 
When the Earth grew through collisions with numerous impactors, these impactors melted part of the Earth's mantle and delivered their metallic iron to the Earth. 
The metallic iron of small impactors (at least up to a few hundred kilometers in size) would have been dispersed as droplets and the resulting rainfall would have led to metal-silicate equilibration during the descent to the metal pond at the base of the magma ocean \citep{Stevenson1990}. 
The iron might have passed through the solid-rich deeper mantle without further equilibration with the ambient mantle.
If the abundance of siderophile elements in the mantle recorded the metal-silicate equilibrium at the base of the magma ocean, the magma ocean depth would have been approximately around 28-40 GPa (700-1,200km deep) \citep[e.g.,][]{LiAgee1996, Righteretal1997}. 
However, this model may be too simplistic because it assumes that the core formation occurred by a single stage process, but the Earth's core must have formed though multiple impacts processes \citep{WadeWood2005, Rubieetal2015}.
Thus, the mantle geochemistry is not likely to have recorded the single impact event.
The concentrations of siderophile elements in the mantle reflect very complex processes of the core formation \citep[e.g.,][]{Stevenson1989, Zimmermanetal1999,  Rubieetal2003, DahlStevenson2010, Shietal2013}.

Analytical and numerical studies suggest that a significant fraction of the mantle would have experienced melting by the Moon-forming impact \citep[e.g.,][]{TonksMelosh1993, Canup2008b, deVriesetal2014}.
A simple estimate can be described as follows; for an impacting body with velocity 10 km/s, the specific kinetic energy carried by the body is $5\times10^7$ J/kg. 
The latent heat of melting is about $1\times10^6$ J/kg (for a mean temperature of 2500K). 
Therefore, a Mars-mass projectile would deliver several times more energy than that needed to melt the entire mantle, assuming that the pre-impact state is near the solidus. 
Of course, this does not indicate that the entire mantle is in fact melted upon impact because the heating is heterogeneous and because part of the impact-induced energy is partitioned to the kinetic energy of the system. 
This makes the important point that the extent of melting must be assessed by considering the budgeting of all delivered energy and not merely by considering the shock heating associated with the primary impact (discussed in \ref{Hugoniot}). In this respect, giant impacts differ from small impacts.

It is often assumed that if there is complete melting at one time, then the preservation of geochemical reservoirs or the retention of volatiles (e.g., He) is not possible. To the extent that this assumption is correct, the geochemical evidence appears to contradict models in which there is complete melting. 
However, we will argue that preserving heterogeneity (i.e., lack of complete mixing) is possible even if most or all of the mantle becomes molten by an impact (Section \ref{model:mix}).

We also consider another problem: even if the mantle heterogeneity survived the Moon-forming impact, it could have been erased in the subsequent evolution.
This could have occurred in two different ways: in the period immediately after the giant impact during the period of ``healing'' (cool-down following a giant impact, leading to a thermal state rather similar to that before the giant impact); or during the very long evolution of the magma ocean and solidifying mantle (time scale of millions to even hundreds of millions of years). This paper is concerned only with the first of these.

We aim to make connections between dynamics and geochemical observations in terms of the Earth's early mantle.
Herein, our main questions are as follows: (1) What is the extent of mantle melting after the Moon-forming impact? (2) If the pre-impact Earth's mantle was heterogeneous, did the giant impact erase the signature? (3) Was the post-giant impact cooling (during the subsequent thousands of years) capable of erasing the pre-giant impact heterogeneity? 
We attempt to answer these questions using the three different models: (a) standard, (b) fast-spinning Earth, and (c) sub-Earths.
Smoothed particle hydrodynamics (SPH) is used for the giant impact simulations (Section \ref{SPH}). 
Two different equations of state are used (Section \ref{EOS}). Furthermore, we investigate the possibility that the Moon-forming impactor and Earth were isotopically similar (Section \ref{radmix}).

\section{Model}
\label{model}
\subsection{Smoothed particle hydrodynamics (SPH)}
\label{SPH}
SPH is a Lagrangian method in which a fluid is represented by numerous particles (grids). 
A particle $i$ has a characteristic size $h_i$, which is the so-called smoothing length.
The mass of the particle $m_i$ is distributed within a sphere of radius 2$h_i$.
Each particle has its own density distribution according to its weighting function (kernel).
The density at a given location is calculated as the sum of the density distributions of nearby particles. 
In a standard SPH, each particle has the same mass, therefore a massive object (e.g. a planet) can be resolved better than a less massive object (e. g., a Moon-forming disk). 
The details of SPH are included in previous studies \citep[e.g.,][]{NakajimaStevenson2014}.
\subsection{Equation of state}
\label{EOS}
We use two different equations of state (EOS) for the mantle materials.
One of the equations represents forsterite Mg$_2$SiO$_4$ (hereafter ``forsterite EOS").
Forsterite is the magnesium end member of the olivine solid solution series and is the major mineral phase in the upper mantle (P$< \sim$14GPa).
The forsterite EOS is derived from the semi-analytic equation of state known as M-ANEOS \citep{ThompsonLauson1972, Melosh2007}.
M-ANEOS has been commonly used for hydrodynamic simulations and it can treat phase changes and co-existing multiple phases.
However, it does not correctly describe the high-pressure properties, especially when the starting material is already in a high-pressure phase such as perovskite MgSiO$_3$ (now more correctly designated bridgmanite).
For this reason, we also use an equation of state that represents MgSiO$_3$ liquid (``MgSiO$_3$ liquid EOS").
Since the Moon-forming impact would induce global melting of the mantle as discussed in Section \ref{mixing}, the liquid EOS is more suitable for this calculation. For comparison, we perform a calculation using an MgSiO$_3$ bridgmanite EOS and discuss the results in Sections \ref{mantlestructure} and \ref{S.brid} (Supplementary materials).
Since there is no ready-to-use MgSiO$_3$ liquid EOS for hydrodynamic simulations, we reconstruct an equation of state based on an analytic expression of the Helmholtz free energy and thermal coefficients derived from the first-principles calculations \citep[e.g.,][]{deKokerStixrude2009}.  
The details are described in Section \ref{S.EOS}.

\subsection{Melting criterion}
\label{melt}
We consider that a region is molten if its entropy gain, which is the entropy difference before and after the impact, exceeds the entropy required for melting $\Delta S_{\rm melt}$. %
A wide range of $\Delta S_{\rm melt}$ has been reported, but here we take a high end-member $\Delta S_{\rm melt} \sim 623$ J/K/kg in the subsequent calculations (MgSiO$_3$,  \citealt{StixrudeKarki2005}).
We also assume that the mantle is close to the solidus before the impact, as discussed in detail in Section \ref{initial}.
It should be noted that since the entropy of melting is a direct measure of the disorder accompanying melting, it should be relatively insensitive to pressure or temperature, unlike the latent heat of melting.

\subsection{Mixing criterion}
\label{model:mix}
As discussed in Section \ref{mantlestructure} in greater detail, one of the most important consequences of a giant impact is that the outermost part of the planet would undergo a greater rise in irreversible entropy than the deeper regions because it would undergo higher shock-heating on average.
This leads to $dS/dr>0$, where $S$ is the entropy and $r$ is the mean radial distance from the center of the Earth to an equipotential surface (the equipotential surfaces within the planet are oblate due to the fast rotation of the planet).
Therefore, the mantle is stable with respect to thermal convection immediately after the giant impact. 
Under this circumstance, even if the mantle is molten, the mantle would not spontaneously convect or homogenize. 
After a giant impact, there is a variation in the entropy rise within the same $r$, but we ignore the azimuthal dependence (discussed in Section \ref{dismix}). 
Our calculations are carried out using entropy values that are averaged over latitude and longitude. 
We discuss the possible shortcomings of this approach in Sections \ref{dismix} and \ref{S.add.Mx}. 

The key issue here is that this entropy profile ($dS/dr>0$) does not necessarily mean that the mantle remains unmixed. 
The giant impact transfers a large amount of kinetic energy to the mantle, which may dynamically mix this stably stratified mantle.
Our SPH calculations dampen the velocity shears on a short timescale (but without allowing turbulence), indicating an unrealistically large viscosity, and therefore they are not suitable for analyzing the mixing directly. 
We must instead use the velocity fields and energy state at an early stage after the giant impact to compute the extent of mixing.  

We investigate the mixing state of the mantle based on an energy balance.
We assume that the mantle is unstable and mixed if the impact-induced kinetic energy of the mantle $\Delta \rm{KE}$ exceeds the potential energy of the stably stratified mantle $\Delta \rm{PE}$ by approximately a factor of two, i.e., $\Delta \rm{PE}/\Delta \rm{KE} < 0.5$ (see Section \ref{S.mixing}). 
The most important source of $\Delta \rm{KE}$ is differential rotation arising from the oblique impact, which is a necessary attribute of any giant impact that is capable of forming the Moon.
$\Delta {\rm KE}$ is calculated as the kinetic energy differences in the mantles between before and after the impact, which is directly obtained from the SPH calculation (here, ``after" indicates a state where the Earth rotates as a rigid body). 
The key point here is that $\Delta {\rm KE}$ is not the same as the kinetic energy of the post-impact Earth if the Earth has an initial spin (this is the case for b). 
This is because the velocity shear is induced by the impact, not by the initial rotation of the planet and it is this differential rotation that is responsible for the instability.
The angular momentum is of course conserved in this process of energy redistribution.
In an overturn that leads to mixing, half of the $\Delta \rm{KE}$ contributes to overcoming gravity, and half is dissipated as heat. Therefore, in this case, the $\Delta \rm{KE}$ is sufficient, by more than a factor of two, to overcome the potential energy of the stably stratified mantle. 

$\Delta {\rm PE}$ is the potential energy difference between the stability stratified mantle ($dS/dr>0$) and mixed mantle ($S(r)=S_{\rm ave}=$const., where $S_{\rm ave}$ is the averaged entropy of the pre-mixing mantle).
$\Delta {\rm PE}$ can be expressed approximately as
\begin{equation}
\Delta PE= \int_{R_{\rm CMB}}^{R_{\rm s}} 4 \pi r^2 V(r) \Delta \rho dr,
\label{eq:potential}
\end{equation}  
where
\begin{equation}
\Delta \rho =\left(\frac{\partial  \rho}{\partial S} \right)_p (S(r)-S_{\rm ave}).
\label{eq:deltarho}
\end{equation}  
$R_{\rm CMB}$ is the distance from the center to the core-mantle boundary (CMB) and $R_{\rm s}$ is the planetary radius.
$V(r)$ is the potential energy per unit mass at $r$.
The gravitational potential of a sphere with the core density $\rho_1$ and mantle density $\rho_2$ is
\begin{equation}
V(r)= -\frac{4\pi G}{3}\frac{\rho_1 R^3_1+\rho_2(r^3-R^3_1)}{r}-2 \pi G \rho_2(R^2_2-r^2),
\label{eq:potential}
\end{equation}  
where $R_1$ and $R_2$ are the core and planetary radii ($R_1<r<R_2$), respectively, and $G$ is the gravitational constant. 
The value of $(\partial \rho/\partial S)_p \equiv - \alpha T \rho/C_p$ depends on the choices of EOS, $V$ and $T$ but it is typically $ -0.3 \sim -0.5$ kg$^2$K/m$^3$J over the parameter range of interest. $\alpha$ is the thermal expansion coefficient, and $C_p$ is the specific heat at constant pressure.

\subsection{Initial conditions}
\label{initial}
Both the target and impactor consist of 70\% mantle (forsterite or MgSiO$_3$ liquid) and 30\% core (iron) by mass. 
We assume that the mantle is compositionally uniform prior to the impact (as discussed in Section \ref{dismix}, it is possible that the mantle was compositionally stratified prior to the giant impact, as would be the case if there were melt migration and a possible basal magma ocean prior to the impact, \citealt{Labrosseetal2007}).
Initially, the mantle has a uniform entropy such that the temperature at the model surface is approximately 2000K (3165 J/K/kg for forsterite and 3350 J/K/kg for MgSiO$_3$ liquid).
It should be noted that this parameter is not important in the model or in the interpretation of the results. 
Of course, the true physical temperature of Earth at its surface could be as low as $\sim$ 300K if any earlier steam atmosphere has collapsed. 
However, the high heat flow in this epoch would lead to temperatures close to the solidus even at a depth of 10km (far too small a distance to be resolved in SPH and a negligible contribution to the total heat capacity). The essential idea here is that the pre-impact Earth cannot cool efficiently at depth to below the solidus temperature (and perhaps not even to the solidus temperature on average) because solid-state convection is unable to eliminate the heat of previous large impacts by convection in the time that has elapsed since the previous giant impact. The most important parameter in assessing the melting upon impact is irreversible entropy production and not temperature.
Here, three types of giant models are investigated: (a) standard, (b) fast-spinning Earth, and (c) sub-Earths.
The parameters considered are the impactor-to-total mass ratio $M_{\rm i}/M_{\rm T}$, the total mass of the target and impactor, $M_{\rm T}$, the scaled impact parameter $b$ ($b=\sin \theta$, here $\theta$ is the impact angle), the impact velocity $v_{\rm imp}$, and the initial spin period, $\tau_{\rm spin}$. The initial conditions of the models are listed on Table \ref{tb:SPHsummary}.
\section{Results}

\label{results}
\subsection{Comparison of equations of state}
\begin{figure*}
  \begin{center}
     \includegraphics[scale=1]{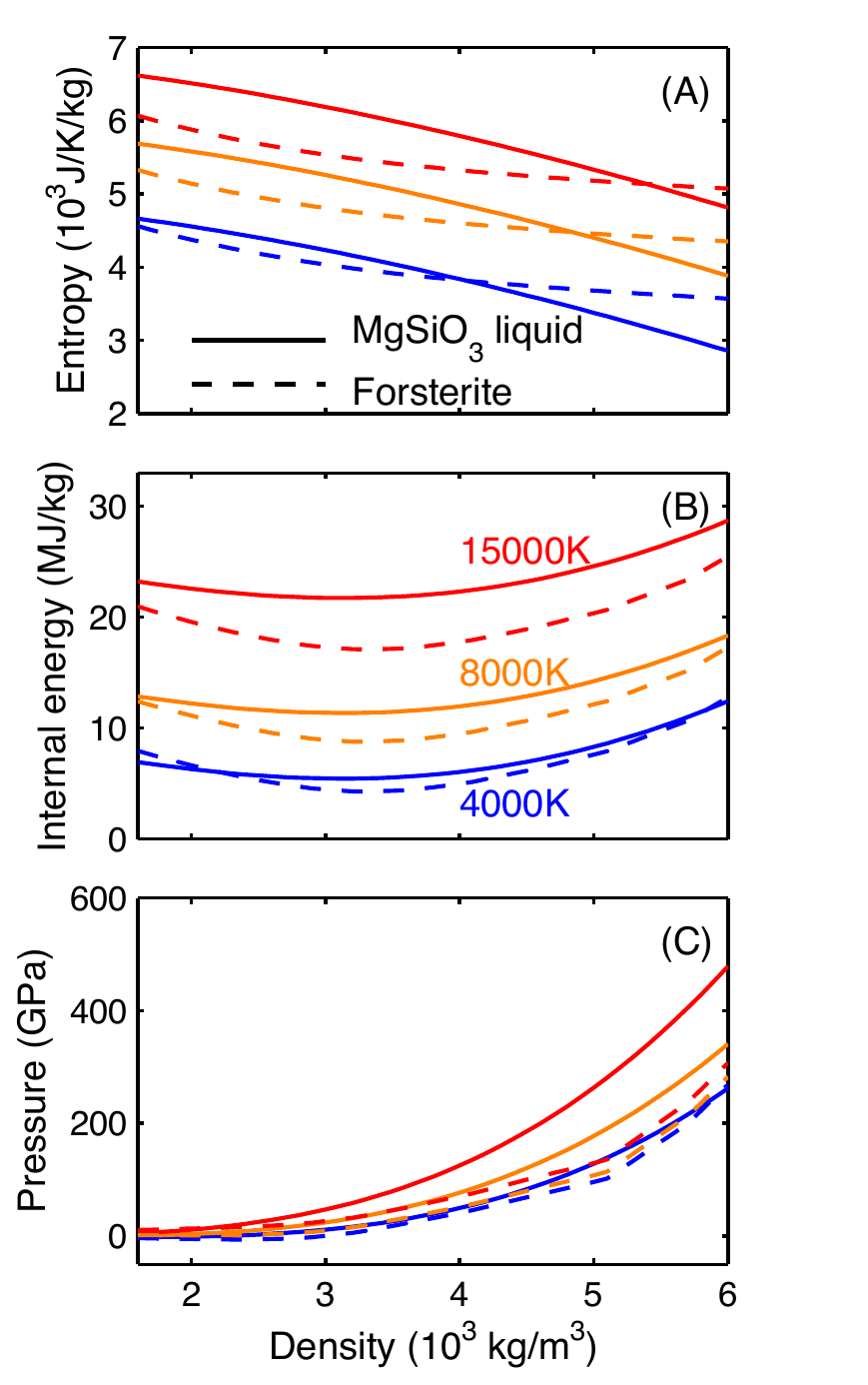}
  \end{center}
  \caption{The two equations of state at various temperatures. The blue, orange and red lines correspond to 4000K, 8000K, and 15000K, respectively. The solid and dashed lines represent the MgSiO$_3$ liquid and forsterite EOS.}
\label{fig:EOS}
\end{figure*}

The behaviors of the two EOS are shown in Figure \ref{fig:EOS}. 
One of the differences is that the entropy of the forsterite EOS is concave-up, whereas that of the MgSiO$_3$ liquid EOS is concave down (Figure \ref{fig:EOS}A).
This disparity can be explained by the signs of $(\partial^2 S/\partial^2 \rho)_{T}$.
For the MgSiO$_3$ liquid EOS, this expression becomes $(q+1)C_V \gamma / \rho^2$, where $\gamma=\gamma_0 (V/V_0)^q$ is the Gr$\ddot{\rm u}$neisen parameter (discussed in Section S1.1) and $C_V$ is the specific heat at a constant volume. 
The exponent $q$ is positive for solids and negative for liquids \citep{Stixrudeetal2009}.
Therefore, the exponent $q$ of the MgSiO$_3$ liquid EOS is chosen to be $-1.6$.  
On the other hand, the forsterite EOS (M-ANEOS) does not differentiate liquids from solids and assumes a positive $q$ ($\gamma=\gamma_0 (V/V_0)+2/3(1-V/V_0)^2$, Equation 4.11, \citealt{ThompsonLauson1972}). 
Therefore, $(\partial^2 S/\partial^2 \rho)_{T}$ is positive in this EOS.
The behaviors of the internal energies are similar (Figure \ref{fig:EOS}B). 
The pressures of the two EOS are not significantly different at low temperature ($T=4000$K) but the difference becomes larger at higher temperatures, which stems from the difference in $q$ (Figure \ref{fig:EOS}C) ($(\partial P/\partial T)_\rho=\gamma C_V/V$ is proportional to $\rho^{1-q}$, derived from Equation 8, Section S1.1). 

\subsection{Mantle structure}
\label{mantlestructure}
\begin{figure*}
  \begin{center}
     \includegraphics[scale=1]{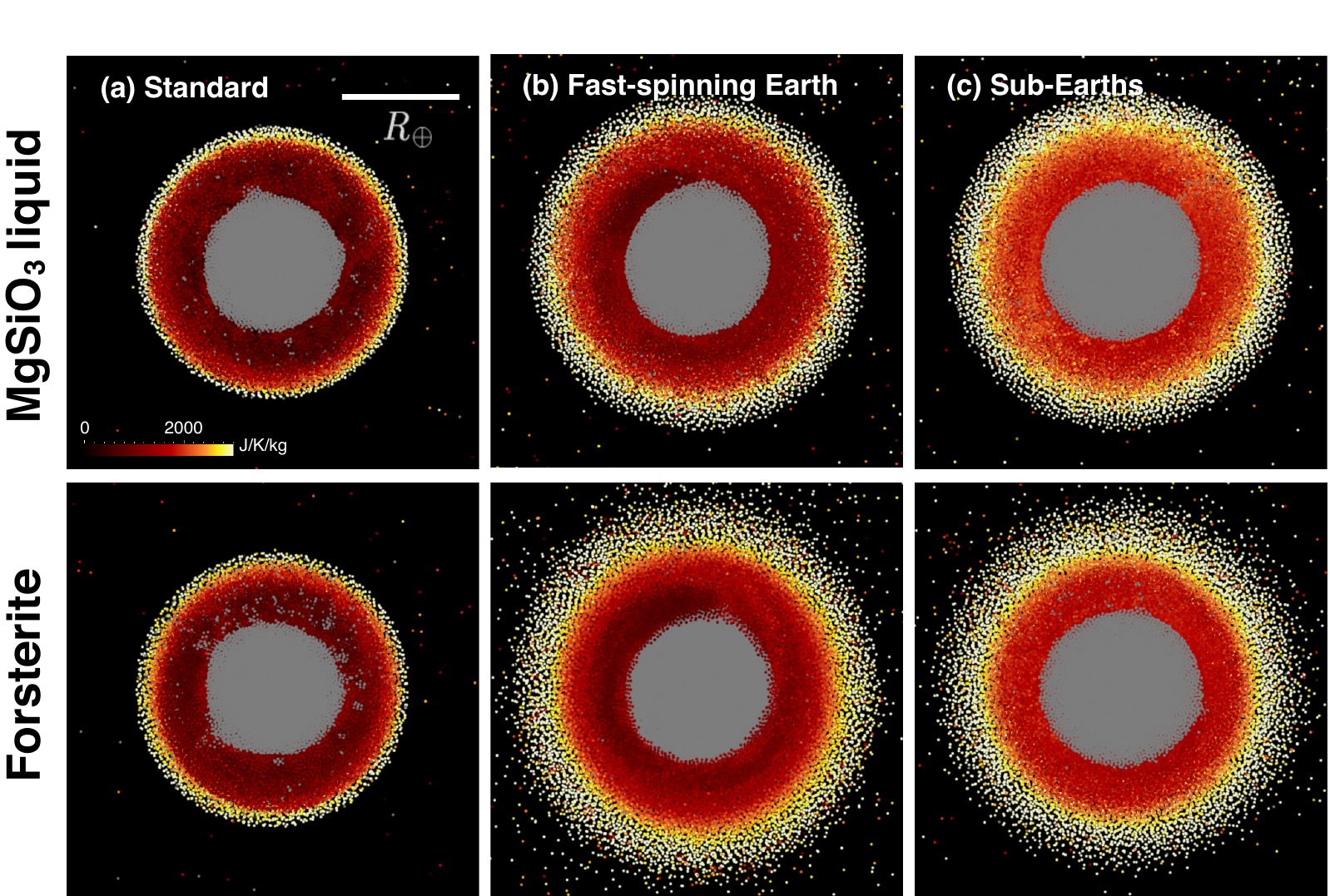}
  \end{center}
  \caption{Cross-section of the mantle after the impact.
  The color gradient scales with the entropy gain of the mantle material. Iron is shown in grey. The top panels show the case with the MgSiO$_3$ liquid EOS and the bottom panels show the case with the forsterite EOS.
 }
\label{fig:mantlesnap}
\end{figure*}

Figure \ref{fig:mantlesnap} shows a cross section of the Earth's mantle after the Earth reaches its equilibrium state ($\sim 1-2$ days after the impact).
The cross-section lies on the $z=0$ plane, which includes the center of the planet, and is perpendicular to the Earth's spin axis ($z$ is parallel to the spin axis).
The color gradient scales with the entropy gain and iron is shown in grey.
As discussed in Section \ref{model:mix}, the entropy is higher near the surface ($dS/dr>0$). Thus, the mantle is thermally stratified (no convection).
In (a), the impactor hits the surface of the Earth twice and becomes disrupted by the impact itself and the tide from the planet.
The surface of the Earth is more shock-heated than the inner part partly because the surface is closer to the impact point and partly because the heavily shock-heated impactor envelops the surface of the Earth.
In (b), the impact velocity is so high that the impactor penetrates all the way to the CMB (Figure \ref{fig:presstime}, b1). 
This energetic impact strips off outer parts of the mantle, which is a desirable feature for producing a disk that is composed of the Earth's mantle rather than projectile.
This strong impact heats the deep mantle materials, which are buoyant and come up to the surface, whereas the ambient colder materials flow to the inner region.
Additionally, the outer part of the Earth's mantle is ejected in the $z$-direction after the impact, which eventually falls back and hits the surface of the Earth.
These materials are highly shock-heated through these processes and eventually distributed on the surface of the Earth.
In (c), the two objects collide each other several times and eventually merge into a single planet.
The mantle becomes highly shock-heated during these processes. 
The surface is more severely shock-heated because it is close to the impact points. 
Thus, although $dS/dr>0$ is a common feature, the reasons for this state differ between the models.

The azimuthal dependence of the entropy is not significant. 
In (a), the impactor hits different parts of the mantle. 
Additionally, the mantle of the impactor, which is highly shock-heated, eventually falls on and covers the surface of the Earth's mantle.
Through these processes,  the Earth's mantle becomes approximately uniformly shock-heated.
In (b), the mantle shows a slightly greater dependence than the other cases. 
The entropy difference at the same $r$ can reach up to a couple of hundred J/K/kg, but it has only minor effects on the outcome (discussed in Section \ref{dismix}).
In (c), the mantle is approximately uniformly heated through the multiple collisions.

\begin{figure*}
  \begin{center}
     \includegraphics[scale=1]{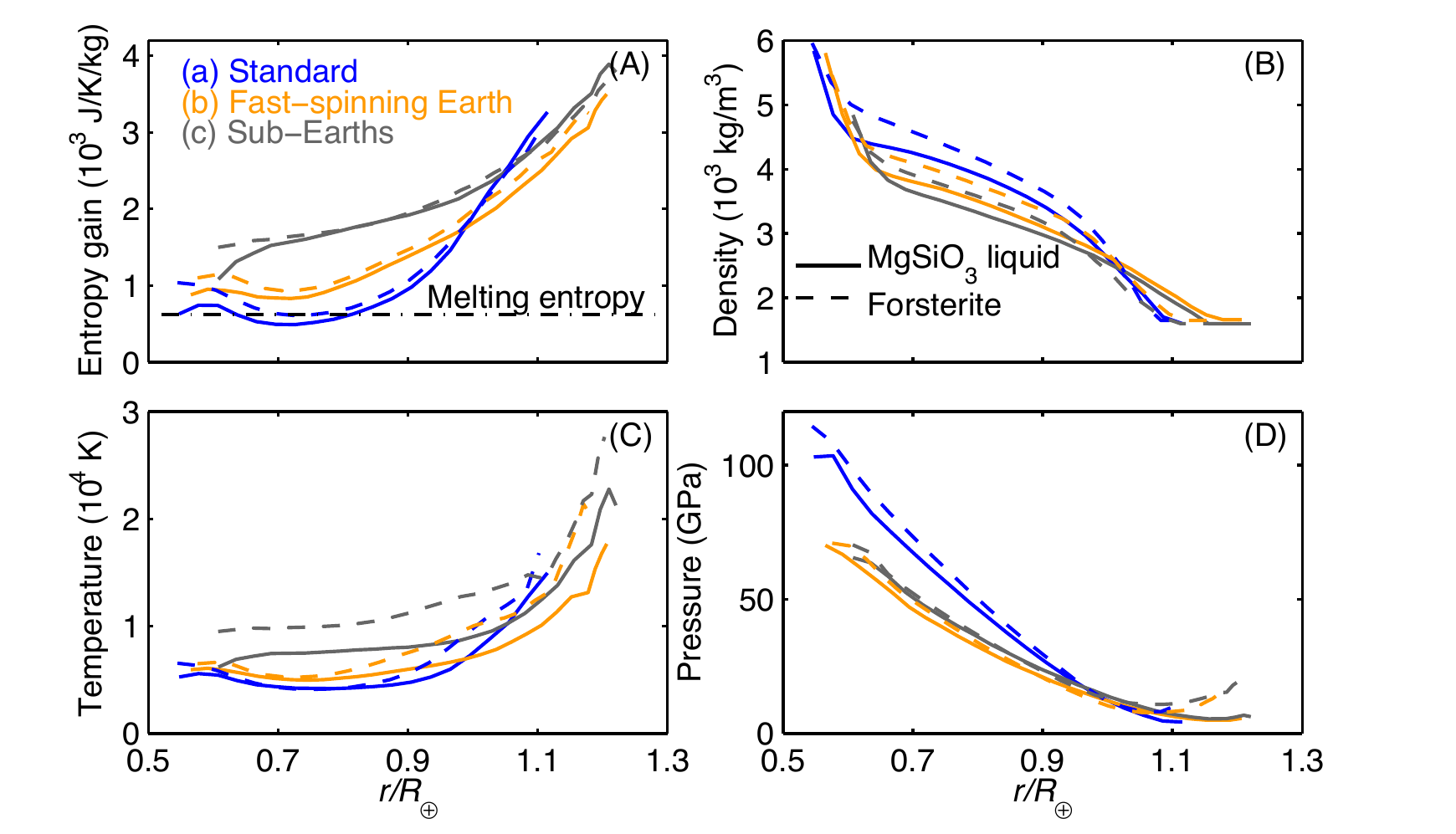}
  \end{center}
  \caption{Mantle structure as a function of $r$. The blue, orange and grey lines represent the standard, fast-spinning and sub-Earth cases. The solid and dashed lines represent the MgSiO$_3$ and forsterite EOS. The dash-dot line represents the entropy of melting. Although the density and temperature are slightly different, the entropy is quite similar for the different EOS. The pressure is also nearly the same except for the outer boundary.
 }
\label{fig:mantleradial}
\end{figure*}

The two EOS provide similar results for each type of impact.
Figure \ref{fig:mantleradial} shows the entropy, density, temperature and pressure of the mantle. 
We set the minimum density $\rho_{\rm min}$ to avoid numerical problems, as described in the Section \ref{outB}.
In Figure \ref{fig:mantleradial}A, $dS/dr>0$ is clearly shown.
Although the MgSiO$_3$ liquid EOS provides a slightly smaller entropy gain than the forsterite EOS, the entropy distribution is very similar between the two EOS (Figure \ref{fig:mantleradial}A).
A clear difference is that in the forsterite EOS, the mantle  is ``puffier" than in the MgSiO$_3$ liquid EOS in (b) and (c) (Figure \ref{fig:mantlesnap}), meaning that the forsterite EOS provides a broader region with small density (Figure \ref{fig:mantleradial}B).
The forsterite EOS has a higher temperature at a given $r$ (Figure \ref{fig:mantleradial}C), simply because the temperature of the forsterite EOS needs to be higher to explain the same entropy as the perovskite EOS in the density range examined (Figure \ref{fig:EOS}A). The pressure distributions are very close until reaching the outermost part (Figure \ref{fig:mantleradial}D).
Nevertheless, the overall physics of the two mantles are not significantly different.
In addition to this, we also perform a simulation for (a) with an MgSiO$_3$ bridgmanite EOS and show that this result is similar to that of the MgSiO$_3$ liquid EOS (Section \ref{S.brid}).

\subsection{Mantle melting and mixing}
\label{mixing}
Figure \ref{fig:mantleradial}A shows that the majority of the mantle experiences melting ($\Delta S_{\rm melt}=623$ J/K/kg).
Specifically, the fractions of the mantle that are melted are (a) 68\%, (b) 99 \%, and (c) 100\% for the MgSiO$_3$ liquid EOS and (a) 80\%, (b)100\%, and (c) 100\% for the forsterite EOS (if $\Delta S_{\rm melt}=500$ J/K/kg, the fractions become (a) 80\%, (b) 100\%, and (c) 100\% for the MgSiO$_3$ liquid EOS and (a) 92\%, (b)100\%, and (c) 100\% for the forsterite EOS. If $\Delta S_{\rm melt}=200$ J/K/kg, the fractions become (a) 98\% for the MgSiO$_3$ liquid EOS and the other cases are 100\%).
Therefore, even if there is no magma ocean prior to the impact, the base of the magma ocean is close to CMB in all cases.
The analysis of the mantle mixing is summarized in Table \ref{tb:mixing}.
The potential energy is normalized by a constant ($0.01 \times \frac{1}{2}$$M_{\oplus}v^2_{\rm esc}$ where $v_{\rm esc}$ is the escape velocity of the Earth).
$\Delta S_{\rm ave}$ represents the average entropy gain due to the impact.
In (a), the ratio $\Delta {\rm PE}/\Delta {\rm KE}$ is much greater than 0.5 in both EOS, which indicates that the mantle may remain unmixed.
In (b), the ratio is still above the critical value, but not as significantly. 
Given the uncertainties in our model, it is difficult to determine the stability of the mantle in this case. This would presumably indicate that the mantle or at least part of the mantle may remain unmixed. 
An important factor here, however, is that once an additional kinetic energy available to mix the mantle is considered, it becomes more likely that the mantle becomes mixed in (b) (discussed in Section \ref{dismix}).
In (c), the ratio is much less than 0.5, which may lead to dynamical mixing of the mantle.
Since $\Delta {\rm KE}$ has larger variations than $\Delta {\rm PE}$ among the models, the determining factor is the kinetic energy.
The kinetic energies of the impacts determined by the different models rank as follows: (a) $<$ (b) $<$ (c) (Table \ref{tb:mixing}).
It should be noted that although the endpoints of the kinetic energy of the mantle in (b) and (c) are similar, $\Delta \rm{KE}$ in (b) is significantly lower than that in (c) because the pre-impact Earth has a large amount of kinetic energy due to the rotation. 
Thus, our study indicates that the standard model (a) is most consistent, the sub-Earths model (c) may be least consistent with the preservation of the mantle heterogeneity, and the fast-spinning Earth (b) may lie between.

\section{Discussion}
\label{discussions}
\subsection{A discrepancy between SPH and the Rankine-Hugoniot equations} 
\label{Hugoniot}
\begin{figure*}
  \begin{center}
     \includegraphics[scale=1]{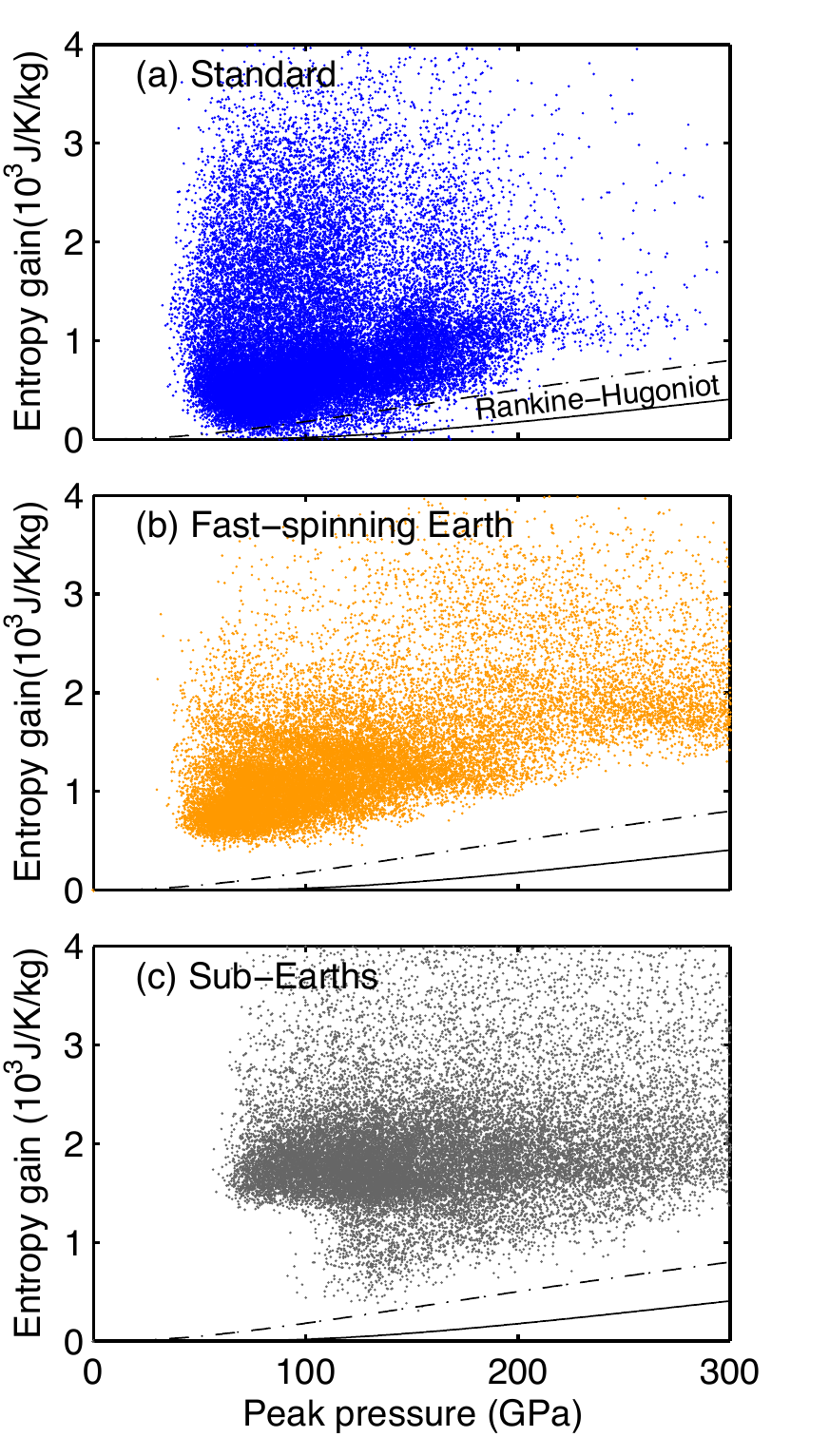}
  \end{center}
  \caption{Entropy gain of each SPH particle in the mantle with the MgSiO$_3$ liquid EOS. The dashed ($p_i=0$ GPa) and solid ($p_i=50$ GPa) lines represent estimated entropy gain based on the Rankine-Hugoniot equations, where $p_i$ is the pre-shock pressure. To examine the lower part of the mantle, this figure only shows SPH particles that were initially under high pressure ($>24$ GPa) before the impact. }
\label{fig:en_hugo}
\end{figure*}

\begin{figure*}
  \begin{center}
     \includegraphics[scale=1]{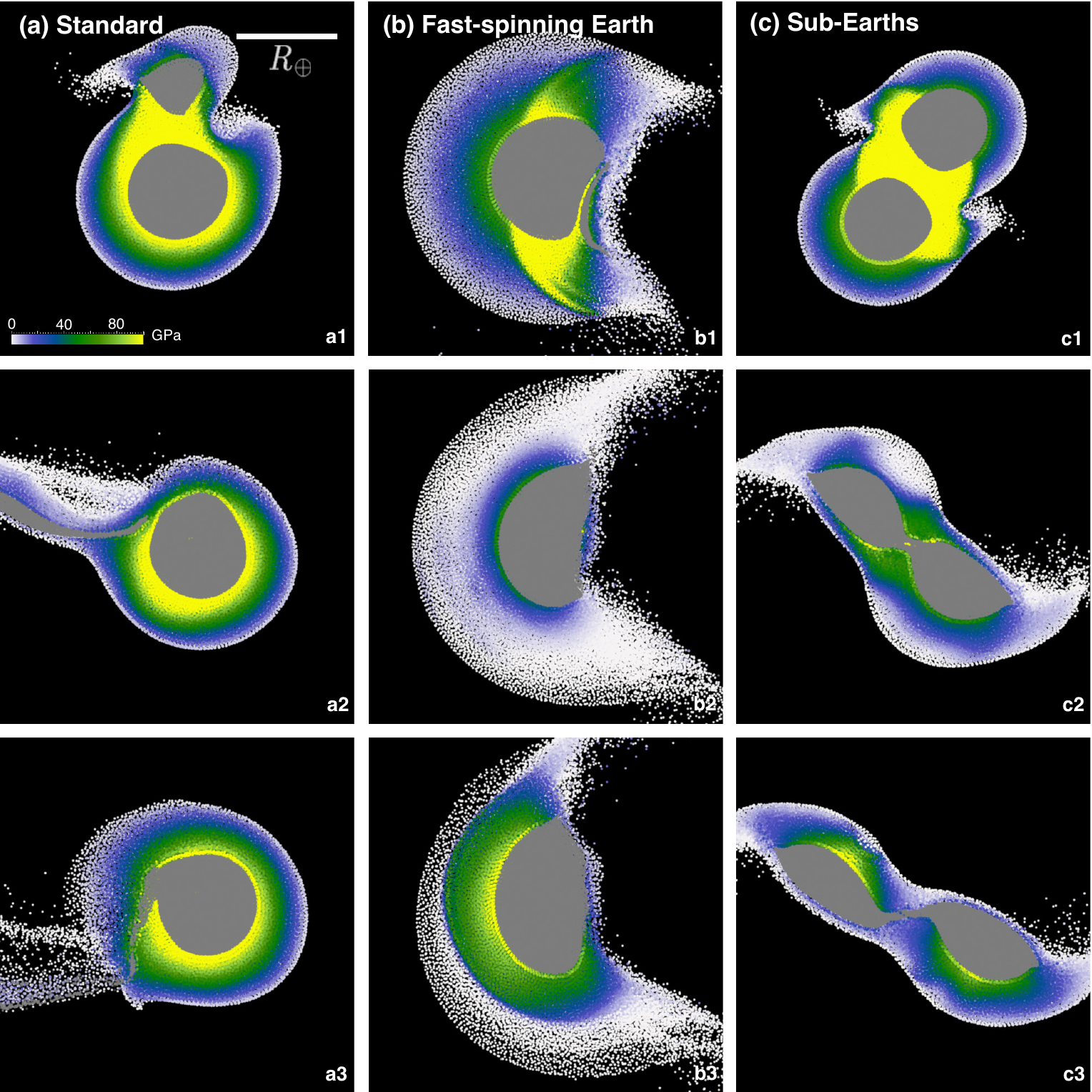}
  \end{center}
  \caption{Snapshots of early stages of the impacts (a1-3: 0.62, 1.0, 1.5 hrs, b1-3: 0.37, 0.60, 0.70 hrs, and c1-3: 0.33, 0.60, 0.72 hrs after the start of the simulation). The color gradient scales with the pressure of the mantle material and iron is shown in grey. The Earth deforms significantly and experiences low pressure in (b) and (c) while the deformation is less intense in (a).}
\label{fig:presstime}
\end{figure*}

The Rankine-Hugoniot equations describe the relations between pre- and post-shock states of materials \citep[e.g.,][]{TonksMelosh1993, Stewartetal2014}.   
The equations can predict the entropy increase of a shocked material by a small impact, but  these equations do not model a giant impact very well. 
This is because the entropy increases by a giant impact not only because of the primary impact-induced shock, which provides the peak shock pressure (as shown in Figures \ref{fig:presstime} a1-c1 and \ref{fig:time_p_S}), but also because of subsequent processes.
Figure \ref{fig:en_hugo} shows a discrepancy between the Rankine-Hugoniot equations and SPH calculations.
It exhibits the entropy of each SPH particle (shown in a dot) after the impact as a function of the shock peak pressure. 
The lines represent predictions of entropy gain using the Rankine-Hugoniot equations (\citealt{Sugitaetal2012}, see Section \ref{S.sugita}). 
It is shown that the entropy gain predicted by SPH is generally much larger than the predictions by the equations, especially in (b) and (c).
This is because the mantle is heated due to a number of additional processes, such as reflected shock waves within the Earth as well as planetary deformation followed by gravitational potential energy release.
The latter effect is clearly shown in Figure \ref{fig:presstime}, which represents the pressure evolution during the giant impact.
In (b) and (c), after the impact, a large fraction of the Earth's mantle experiences an extensive expansion and becomes subject to low pressure (b2 and c2).
Subsequently, the part falls back towards the Earth's core and gains a high pressure (b3 and c3).
Through these processes, the part of the mantle efficiently gains entropy.  
In (a), the extent of such deformation is not as prominent as the other cases; therefore, its entropy increase is closer to the analytical estimates.
The time dependent relationship between the pressure and entropy is discussed further in Section \ref{S.PressEn}.
Thus, the Rankine-Hugoniot equations of state predict only a part of the entropy gain (due to the primary impact-induced shock); therefore they do not represent the total entropy gain by a giant impact.

\subsection{Mixing analysis}
\label{dismix}
For this mixing analysis, we assume that $\Delta {\rm KE}$ is the kinetic energy of the differential rotation of the mantle. 
However, differential rotation is not the only source of energy available for mixing. In addition to flows produced by the impact itself, there will be excitation of normal modes of the planet (we have observed these in longer runs of SPH simulations for close encounters or glancing impacts). These are analogous to waves on an ocean and have an energy that oscillates between gravitational energy and kinetic energy of the fluid motion, with equipartition of the average kinetic energy and average gravitational energy. The mean kinetic energy is therefore simply related to the gravitational energy for a distortion of an equipotential surface or the planet as a whole by a vertical distance $h$. These normal modes will ``ring down'' slowly, but in a region of static stability the distortion of equipotential surfaces creates horizontal density gradients that drive rapid baroclinic instabilities. 
Thus, some significant fraction of this energy (the precise amount is not easily estimated) can go into mixing.
Here we make a simple estimate for the additional kinetic energy for mixing: the gravitational energy by distortion for a uniform density sphere is proportional to $\frac{GM_{\oplus}^2}{R_{\oplus}}(\frac{h}{R_{\oplus}})^2$. Approximately, $\frac{h}{R}_{\oplus}$ is $\sim 0.1$ in (a) and $> 0.3$ in (b) and (c). Thus, the gravitational energy becomes $\sim$ 1, 10 and 10 (in 0.01 $\times \frac{1}{2}M_{\oplus}v_{\rm esc}^2$, which is the unit of $\Delta {\rm KE}$ in Table \ref{tb:mixing}). 
Assuming that approximately half of this energy eventually goes into heat and the other half is available to mix the core (this increases $\Delta {\rm KE}$), corrected $\Delta {\rm PE}/\Delta {\rm KE}$ become 1.60, 0.35, and 0.17 in (a), (b) and (c) (MgSiO$_3$ liquid EOS). Thus, the mantle would still remain unmixed in (a), but the mantle would be mixed in (b) and (c).

We ignore the azimuthal dependence, which is likely a reasonable approximation.
To demonstrate that the approximation is feasible, consider the worst case, the mantle in (b), whose entropy distribution shows the strongest azimuthal dependence. We evenly divide the mantle into four sections according to its azimuthal angle. We observe that the average $\Delta \rm{PE}$ in each section differs from the globally averaged value by less than $10 \%$, which is not sufficient to alter the mixing status.
Another potential concern is that  we only perform one simulation for each model and EOS, therefore it is possible that the ratio $\Delta \rm{PE}/\Delta \rm{KE}$ varies among models, but we show that the ratio we derived is likely to capture the general trend (discussed in Section \ref{S.add.Mx}).

It should be noted that the mixing analysis described in Section \ref{mixing} only determines the global mixing state. 
Therefore, it may be possible to preserve a local heterogeneity even at $\Delta {\rm PE}/\Delta {\rm KE}<0.5$, especially if the ratio is close to the critical value. 
Unfortunately, our SPH does not provide information about local mixing.
To investigate such local mixing, it would be necessary to perform a simulation with much higher resolution, but doing so would be computationally quite expensive. 
Additionally, we do not consider the possibility that the mantle was compositionally stratified before the impact (e.g. a denser layer at the base of the mantle), but if this was the case, it could increase the stability of the mantle. Moreover, we discuss issues and implications of this mixing analysis in Section \ref{S.add.Mx}.

\subsection{Mixing during subsequent cooling}

We turn now to the question of whether the mantle will mix during the post-giant impact cooling. This is a simple problem, conceptually, and it does not require extensive numerical analysis. As the outer regions of Earth cool, the mantle will evolve into two regions, a nearly isentropic region that can convect and a deeper region where the entropy is still lower than that appropriate for the effective radiating temperature of Earth at that time. The deeper region that is of lower entropy is unable to convect since it is still part of the stably stratified zone. Thus, the convective zone propagates downwards as the planet cools and the deepest part of the mantle is only able to convect when the entire region above has cooled to the entropy that this deepest part had immediately after the giant impact. But as we have seen, this entropy is not much larger than it was before the giant impact, at least in the case of the canonical giant impact. This takes a long time relative to the time it takes to make the Moon, as we now demonstrate. Let $T_e$ be Earth's effective temperature for the thermal state corresponding to the isentrope that is only slightly higher (and hotter) than the pre-giant impact state. At that epoch, the cooling equation takes the form

\begin{equation}
4\pi R^2\sigma T_e^4=-\frac{d}{dt}(MC_V T_{\rm m})
\label{eq:mixing1}
\end{equation}  	
where $M$ is the mantle mass and $T_{\rm m}$ is the mean temperature at that epoch. Accordingly, the time $\tau$ that takes to cool an amount $\Delta T_{\rm m}$ is
\begin{equation}
\tau \sim 5 \times 10^3  \left(\frac{\Delta T_{\rm m}}{1000 {\rm K}}\right) \left(\frac{1000 {\rm K}}{T_e}\right)^4 \rm{yrs}
\label{eq:mixing2}
\end{equation} 	

This is an underestimate because we assume a typical specific heat rather than the additional term that comes from freezing. Importantly, $T_e$ can be low at this point, though its exact value depends on whether there is a steam atmosphere. For example, at $T_e$= 500K, this cooling time for 1000K is tens to hundreds of thousands of years, which is much longer than the estimated time for making the moon (hundreds to thousands of years). The reason for this is obvious: Earth's mantle has a much greater heat capacity than the Moon-forming disk and yet a smaller area from which to cool. This is the time that must elapse before any mixing of the deepest mantle material can even begin. Even at that time, mixing will not occur if there is a basal magma ocean prior to the impact. It is often assumed in the literature that mixing occurs if there is liquid but it is important to stress that the state of the material is not the crucial issue here. Counterintuitive though it may seem, it is harder to mix a stably stratified liquid than to mix a solid of the same stratification because the thermal anomalies associated with convection in a liquid are much smaller than those for solid state convection.

\subsection{Impactor's isotopic signature}
\label{radmix}
For the standard scenario (a), the impactor needs to have isotopic signatures similar to those of the Earth to explain the geochemical evidence.
To gain a better idea of how likely this similarity could be in the Grand Tack model, we run similar analyses to those of \cite{PahlevanStevenson2007}. Although \cite{KaibCowan2015} have conducted isotopic analyses of a model that is similar to the Grand Tack, the initial conditions are more simplified than the original model.

We take 30 planets from eight simulations from \cite{Walshetal2011} (each simulation produces three to four terrestrial planets) and examine the difference in the oxygen isotope ratios between the target and impactor.
We assume that the oxygen isotopic ratio in the protoplanetary disk linearly varies as a function of the heliocentric distance $r'$, $\Delta^{17} {\rm O}(r')=c_1 \times r' +c_2$. 
Two pairs of $c_1$ and $c_2$ are used; case 1: $c_1$ and $c_2$ are chosen such that $\Delta^{17} {\rm O}(r')=0$ $\text{\textperthousand}$ at 1 AU and  $\Delta^{17} {\rm O}(r')=0.321$ $\text{\textperthousand}$ at 1.5 AU($\Delta^{17}{\rm O}$ of Mars), and case 2: $c_1$ and $c_2$ are chosen such that the Earth and Mars analogues in each simulation have $\Delta^{17} {\rm O}(r')=0$ $\text{\textperthousand}$ and $0.321$ $\text{\textperthousand}$, respectively (this is the same setting as those of \citealt{PahlevanStevenson2007}). Simulations that do not produce Mars analogues are discarded.
Figure \ref{fig:GT} shows histograms of the differences in the oxygen isotopic ratios.
The top and bottom planes correspond to case 1 and case 2.
The left panels include all of the impacts and the right panels focus on the last giant impact for each planet.
Even if we take the stricter criterion (the isotopic difference should be less than $0.005$ $\text{\textperthousand}$), there is one last giant impact out of 30 events and one out of 19 events that satisfy this criterion in case 1 and case 2, respectively (Figure \ref{fig:GT} right panel). 
It should be noted that the mass involved in these particular impacts are a couple of Mars masses, so that these do not mimic the Moon forming impact.
Nevertheless, this simple analysis may indicate that the Grand Tack model could produce an impactor that has a similar composition to the Earth and its possibility is around 3-5 $\%$. 
Needless to say, a much larger dataset is needed to estimate the probability. Additionally, other isotopic ratios, such as tungsten and silicon, need to be explained in addition to oxygen (Section \ref{intro}).

\section{Conclusion}
\label{conclusions}
To investigate the relationship between the Moon-forming model and geochemical evidence, we have investigated the initial state of the Earth's mantle after the Moon-forming impact.
Giant impact simulations using the SPH method are performed with the MgSiO$_3$ liquid and forsterite equations of state.
Three impact models are considered here: (a) standard, (b) fast-spinning Earth, and (c) sub-Earths.
We find that the mantle becomes mostly molten after the Moon-forming impact in all of the cases.
It is also shown that the impact-induced entropy gain of the mantle cannot be predicted well by the Rankine-Hugoniot equations.
Based on our analysis on the energy balance of the mantle, we find that the Earth's mantle is likely to stay unmixed in (a), while it may be mixed in (b) and (c).
The extent of mixing is most extensive in (c). 
This is primarily because the impact-induced kinetic energy in (c) is much larger than that of (a) while (b) falls between.
Thus, the standard model (a) is most consistent, the sub-Earths model (c) may be least consistent with the preservation of the mantle heterogeneity, and the fast-spinning Earth (b) may lie between. 
We also find that if the impact does not mix the mantle, the Earth's mantle would remain partially unmixed for more than the Moon accretion time scale ($\sim 100-1000$ yrs).
It is therefore possible, at least in the standard giant impact case (and possibly some others), to avoid complete mixing during and immediately after a giant impact. The issue of subsequent mixing ($> \sim 1000$ yrs) is not a simple problem but is no different from the usual assessment of mantle convection mixing \citep[e.g.,][]{Tackley2012}.

\section*{Acknowledgement}
\label{acknowledgement}
This work is supported by NASA Headquarters under the NASA Earth and Space Science Fellowship Program - Grant NNX14AP26H.
We would like to thank David Rubie and an anonymous reviewer for insightful comments, Kevin Walsh for providing the Grand Tack simulations, Paul Asimow, Kaveh Pahlevan, Aaron Wolf, Sarah Stewart and Tobias Bischoff for helpful discussions, Takaaki Takeda for providing a visualization software, Zindaiji 3.
Numerical computations were partly carried out on GRAPE system at Center for Computational Astrophysics, National Astronomical Observatory of Japan.


\begin{center}
\begin{table*}[ht]
{\small
\hfill{}
\begin{tabular}{|l|c c c c c|}
\hline
\textbf{}& $M_{\rm i}/M_{\rm T}$ & $M_{\rm T}$ & $b$ & $v_{\rm imp}$&$\tau_{\rm spin}$   \\
\hline
(a) Standard &0.13&1.02&0.75&1.0&0\\
(b) Fast-spinning Earth &0.045&1.05&-0.3&20 (km/s)&2.3\\  
(c) Sub-Earths &0.45&1.10&0.55&1.17&0\\ 
\hline
\end{tabular}}
\hfill{}
\caption{The initial conditions of the impact. $M_{\rm i}/M_{\rm T}$ is the impactor-to-total mass ratio, $M_{\rm T}$ is the total mass scaled by the mass of the Earth, $b$ is the scaled impact parameter, $v_{\rm imp}$ is the impact velocity scaled by the escape velocity (except b), and $\tau_{\rm spin}$ is the initial spin period of the target (hrs). }
\label{tb:SPHsummary}
\end{table*}
\end{center}

\begin{center}
\begin{table*}[ht]
{\small
\hfill{}
\begin{tabular}{|l|c c c c |}
\hline
\textbf{}& $\Delta {\rm KE}$ & $\Delta {\rm PE}$ & $\Delta {\rm PE}/\Delta {\rm KE}$ &$\Delta S_{\rm ave}$ \\
\hline
(a1) Standard, liq &1.38&3.01&2.18&1181 \\ 
(a2) Standard, for &1.51&2.82&1.86&1220\\
(b1) Fast-spinning Earth, liq &3.08&2.79&0.91&1645\\
(b2) Fast-spinning Earth, for &3.36&2.45&0.73&1675\\
(c1) Sub-Earths, liq &5.85&1.85&0.32&2210\\
(c2) Sub-Earths, for &5.50&1.26&0.23&2196\\
\hline
\end{tabular}}
\hfill{}
\caption{The outcomes of the impact. $\Delta {\rm KE}$ is the change in kinetic energy (normalized  by $0.01\times \frac{1}{2}M_\oplus v^2_{\rm esc}$, where $v_{\rm esc}$ is the escape velocity of the Earth) and $\Delta {\rm PE}$ is the potential energy required to mix the mantle. $\Delta S_{\rm ave}$ is the average entropy gain of the mantle. liq represents the MgSiO$_3$ liquid and for represents forsterite EOS.}
\label{tb:mixing}
\end{table*}
\end{center}


\begin{figure*}
  \begin{center}
     \includegraphics[scale=1]{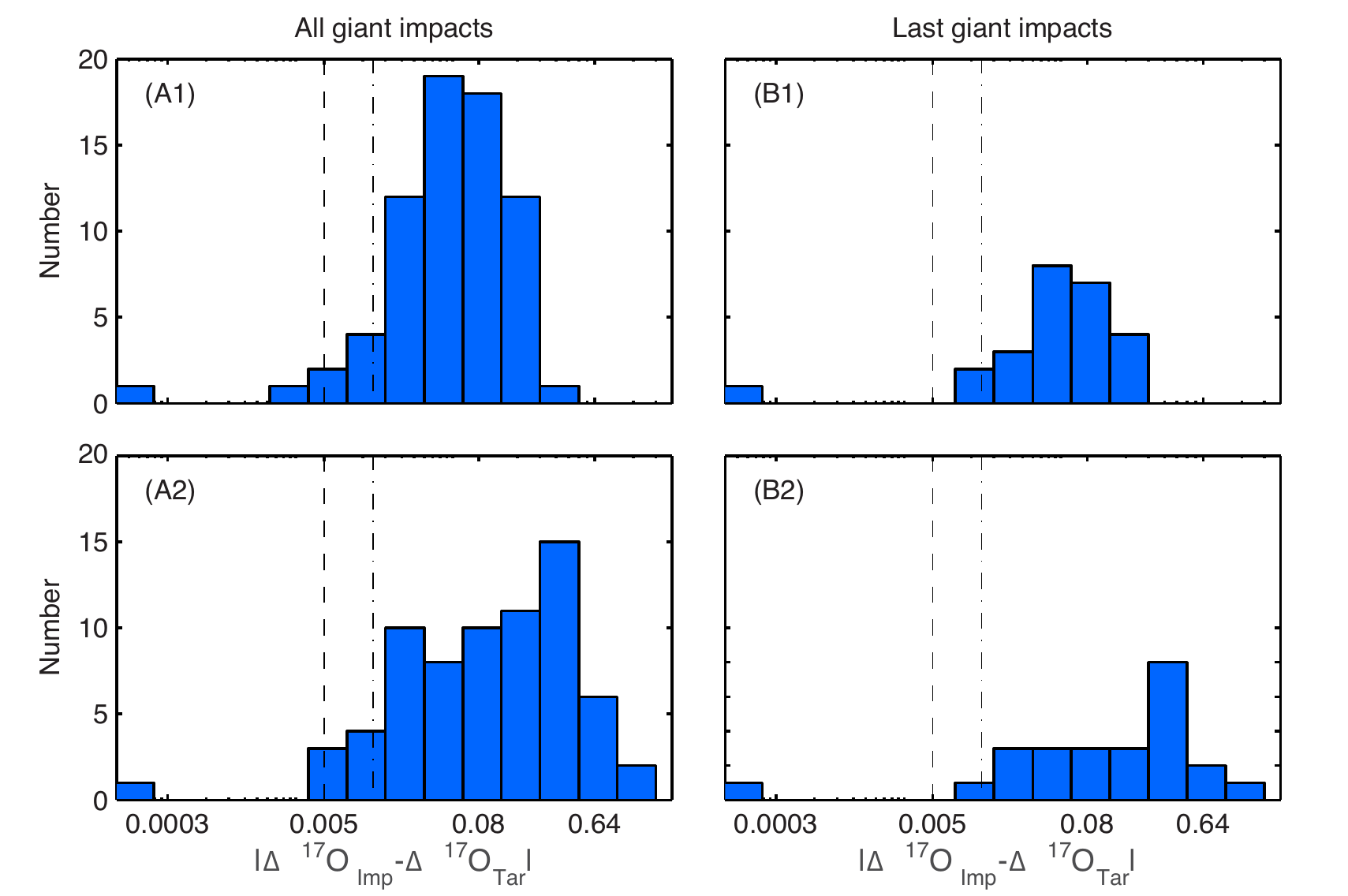}
  \end{center}
  \caption{Histogram of the differences in the oxygen isotopic ratios between the target (proto-Earth) and impactor. The left panel includes all of the giant impacts and the right panel includes only the last giant impact for a planet. The top and bottom panels represent different initial conditions (case 1 and case 2, Section \ref{radmix}).  
The dashed and dash-dot lines indicate the two estimates of the oxygen isotopic differences between the Earth and Moon (0.005$\text{\textperthousand}$ and 0.012$\text{\textperthousand}$, respectively, \citealt{Wiecheretal2001, Herwartzetal2014}).
 }
\label{fig:GT}
\end{figure*}




\newpage

\setcounter{section}{0}


\renewcommand{\thesection}{S\arabic{section}}

\setcounter{equation}{0}
\setcounter{table}{0}



\section{Supplementary Materials}
\subsection{Equation of state}
\label{S.EOS}
Our equation of states for the MgSiO$_3$ liquid follow the formulae previously derived \citep[e.g.,][]{deKokerStixrude2009, Stixrudeetal2009}.
The Helmholtz free energy is written as
\begin{equation}
F(V,T)=F_0+F_{\rm cmp} (V,T_0)+F_{\rm th}(V,T),
\label{eq:stix_F}
\end{equation} 
where $F_0=F(V_0, T_0)$ is the free energy at the reference volume, $V_0$, and temperature, $T_0$. $F_{\rm cpm}(V,T_0)$ and $F_{\rm th}(V,T)$ are the compressional and thermal contributions to the free energy, respectively. $F_{\rm cmp}$ is
\begin{equation}
F_{\rm cmp}=9K_{T0}V_0\left(\frac{1}{2}f^2+\frac{1}{6}a_3 f^3 \right),
\label{eq:Fcmp}
\end{equation}
where 
\begin{equation}
a_3=3(K'_{T0}-4),
\label{eq:a3}
\end{equation} 
\begin{equation}
f=\frac{1}{2} \left[ (V_0/V)^{2/3}-1\right].
\label{eq:f}
\end{equation} 
$K_{T0}$ is the isothermal bulk modulus (at $T=T_0$), and $K'_{\rm T0}$ is its pressure derivative at $p=0$ and $T=T_0$. $F_{\rm th}$ is written as 

\begin{equation}
F_{\rm th}=-\int^{T}_{T_0} S(V,T')dT'.
\label{eq:Fcmp}
\end{equation}
The entropy $S(V,T) $is described as
\begin{align}
S(V,T) & =S_0+\int_{V_0}^V \frac{C_V \gamma (V',T_0)}{V}dV'   \nonumber \\
& + \int_{T_0}^{T}\frac{C_V(V,T')}{T'}dT',
\label{eq:en}
\end{align}
where $S_0=S(V_0,T_0)$, $C_V$ is the specific heat that is assumed to be a constant. $\gamma$ is the Gr$\ddot{\rm u}$neisen parameter, which is described as
\begin{equation}
\gamma=\gamma_0 \left( \frac{V}{V_0}\right)^q,
\label{eq:a3}
\end{equation} 
where, $\gamma_0$ and $q$ are constants. 

The internal energy, $E$, and pressure, $P$, are described as
\begin{align}
E(V,T) & =E_0+9K_{T0} V_0 \left( \frac{1}{2}f^2+\frac{1}{6}a_3 f^3  \right)  \nonumber \\
& +C_V (T-T_0)+C_V T_0 \int_{V_0}^{V} \frac{\gamma(V',T_0)}{V'}dV',
\label{eq:ene}
\end{align}
\begin{align}
P(V,T) & =3 K_{T0} (1+2f)^{5/2}(f+\frac{a_3}{2}f^2) \nonumber \\
&+C_V (T-T_0) \frac{\gamma(V,T_0)}{V}.
\label{eq:press}
\end{align}
Here, $E_0=E(V_0,T_0)$. $\rho_0=(1/V_0)$, $T_0$, $K_{T0}$, $K'_{T0}$, $C_V$, $\gamma_0$, $q$, $E_0$, and $S_0$ are listed in Table \ref{tb:tb1}.

\begin{center}
\begin{table*}[ht]
\setcounter{table}{0}
\renewcommand{\thetable}{S\arabic{table}}
{\small
\hfill{}
\begin{tabular}{|c c  c c c c c c c |}
\hline
$\rho_0$(kg/m$^3$)&$T_0(K)$&$K_{T0}$(GPa)&$K'_{T0}$&$C_V$(J/K/kg)&$\gamma_0$&$q$ &$E_0$(MJ/kg)& $S_0$(kJ/K/kg) \\
\hline
2650&2000&27.3&5.71&1480&0.6&-1.6 & 2.64& 3.33\\  
\hline
\end{tabular}}
\hfill{}
\caption{Constants for the MgSiO$_3$ liquid EOS.}
\label{tb:tb1}
\end{table*}
\end{center}

\subsection{Mixing criterion}
\label{S.mixing}
In a simple shear flow, the criterion for a Kelvin-Helmohltz instability (effectively the criterion for mixing) is Ri$\equiv N^2/(du/dz)^2 <1/4$. 
$Ri$ is the Richardson number of the system, $N$ is the Brunt-Vaisala frequency, $N^2 \equiv -g(d\rho/dz)/\rho$, $u$ is the velocity, $z$ is the direction perpendicular to the flow, $g$ is the gravity, and $\rho$ is the density. 
Ri is related to the ratio of potential energy to kinetic energy \citep[e.g.,][]{Taylor1931, Chandrasekhar1961} and is normally defined in terms of fluids with well defined constant density differences or a density gradient, but our system has variable values of all the input parameters. Therefore, we must necessarily restate the problem in terms of the energy budgets rather than explicitly in terms of velocity shear. 
The kinetic energy difference $\Delta$KE per unit area for a layer of thickness $L$ between the initial state (with shear) and the final state (with no shear but the same linear momentum) is 
\begin{equation}
\frac{1}{2}\overline{\rho} \int^{L/2}_{-L/2} (u(z)^2-(u_0/2)^2)dz=L \overline{\rho} u_0^2/24,
\label{ap2_ke}
\end{equation} 
where $u_0$ is the initial velocity difference between the top and bottom and $\overline{\rho}$ is the mean density. 
The gravitational potential energy difference ($\Delta$PE) between initial and final (fully mixed) states is 
\begin{equation}
 \int^{L/2}_{-L/2} ( \overline{\rho}-\rho(z))g z dz= \rho N^2L^3/12,
\label{ap2_pe}
\end{equation} 
where the density variation is assumed small. Therefore, the Richardson number criterion becomes $\Delta$KE$>$2$\Delta$PE and the physical interpretation is that one must provide not only the energy to overcome the potential energy difference but also the energy to mix (which shows up as heat from the dissipation of small scale turbulent motions). In the analysis provided by \cite{Chandrasekhar1961} (p. 491), the $\Delta$PE he defines is for complete overturn (that is, the new density profile is the exact opposite of the initial density profile), which is a different setting from ours. Therefore, his criterion differs from ours (mixed if $\Delta {\rm KE}>\Delta {\rm PE}$ in his anaysis). 

\subsection{MgSiO$_3$ bridgmanite EOS}
\label{S.brid}
Figure \ref{fig:brid} shows cross-sections of the mantles after the impact with the MgSiO$_3$ liquid and MgSiO$_3$ bridgmanite EOS.
The thermodynamic parameters for the bridgmanite EOS is listed in Table \ref{tb:brid}.
The entropy gains are slightly different, but the extent of shock-heating and the feature of $dS/dr>0$ are similar among these cases. 
One might expect that a liquid mantle may gain higher entropy than a solid one based on  the study done by \cite{Karato2014} that suggests that the surface of a molten mantle gains higher entropy by impact than a solid surface due to its smaller sound speed $C_0$ and negative $q$ for the liquid (for the definitions of these parameters, see Section \ref{S.sugita}). 
However, the difference of the two EOS in $C_0$ becomes smaller at a greater depth. This would diminish the difference in the entropy gain among the two EOS.
In addition, \cite{Karato2014} use the Rankine-Hugoniot equations to describe the physics of the planetary surface, but we cannot use these equations to predict the entropy gain of the entire mantle as discussed in Section 4.1.

\begin{figure*}
\setcounter{figure}{0}
\renewcommand{\thefigure}{S\arabic{figure}}
  \begin{center}
     \includegraphics[scale=1]{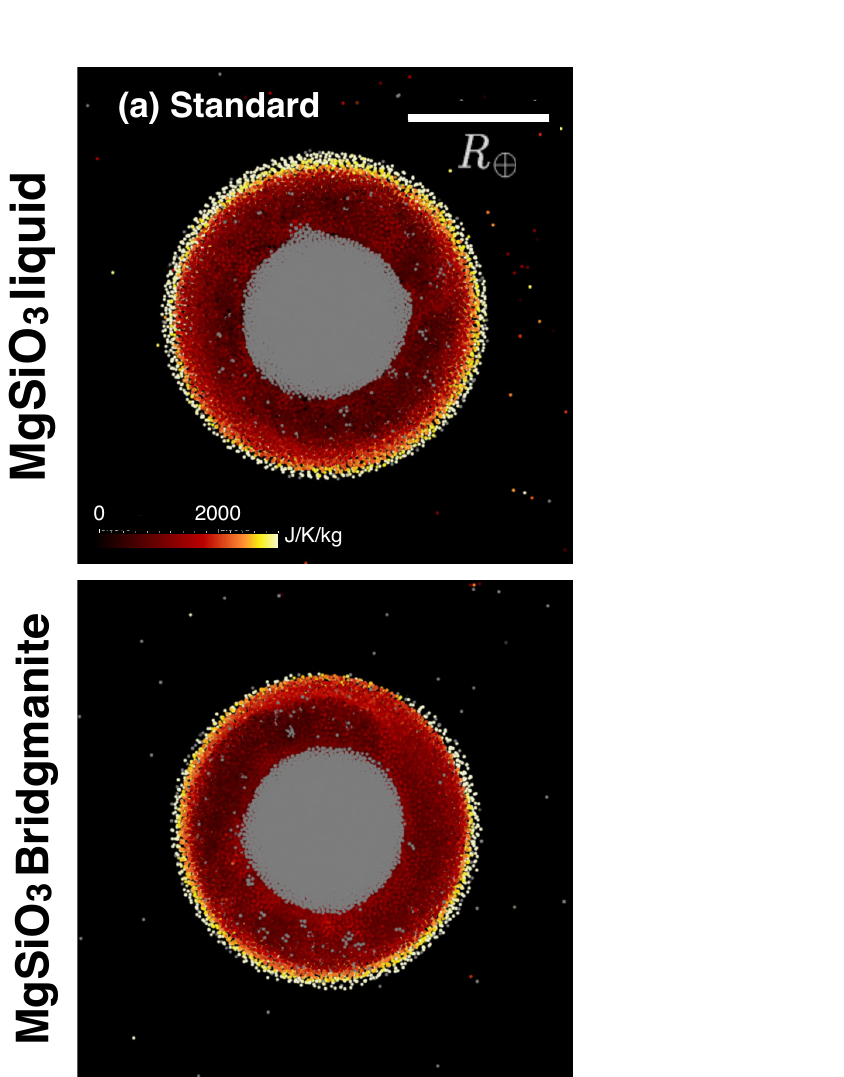}
  \end{center}
  \caption{Entropy of the Earth's mantle after the impact. The top panel shows the case with the MgSiO$_3$ liquid EOS (the same as Figure 2 in the main text) and the bottom panel shows the case with the MgSiO$_3$ bridgmanite EOS.}
\label{fig:brid}
\end{figure*}

\begin{center}
\begin{table*}[ht]
\setcounter{table}{1}
\renewcommand{\thetable}{S\arabic{table}}
{\small
\hfill{}
\begin{tabular}{|c c  c c c c c c c |}
\hline
$\rho_0$(kg/m$^3$)&$T_0(K)$&$K_{T0}$(GPa)&$K'_{T0}$&$C_V$(J/K/kg)&$\gamma_0$&$q$ &$E_0$(MJ/kg)& $S_0$(kJ/K/kg) \\
\hline
3680&2000&200&4.14&1200&1.0&1.0 & 1.995& 2.63\\  
\hline
\end{tabular}}
\hfill{}
\caption{Constants for the equation of state of MgSiO$_3$ bridgmanite.}
\label{tb:brid}
\end{table*}
\end{center}

\subsection{Correction of the outer boundary} 
\label{outB}
The density of the outer edge of the mantle is corrected because the simulation itself does not provide an accurate value.
One of the reasons is that the standard SPH cannot describe a large density difference (e.g., CMB or planet-space boundary).
The density of a particle at the outer boundary becomes too small because the particle does not have many nearby particles; thus, the smoothing length becomes large. 
This leads to a problem that $dP/dr$ at the outermost part of the mantle becomes nearly $0$ or it even becomes positive in (b) and (c) (Figure 3D). 
This state is not physically sensible because the hydrostatic equation is not correctly solved in the region. 

To avoid this numerical problem, we define the minimum density $\rho_{\rm min}=\rho(r_{\rm max})$.
Here, $r_{\rm max}$ is the maximum $r$ whose region satisfies $dP/dr<0$.
If the density at $r>r_{\rm max}$ is lower, the $r$ is recalculated by setting $\rho=\rho_{\rm min}$ and conserving the mass.
Typically, $\rho_{\rm min}\sim 1500 - 1600$kg/m$^3$ (Figure 3B). 
This is uncertain but may be reasonable because this is close to a rough estimate of the density at such a high temperature.
The density at the outer edge can be approximated as $\rho \sim  \rho_0 (1-\alpha T) \sim 1577 $kg/m$^3$ at $\alpha=2.7 \times10^{-5}$ \citep{Fiquetetal2000}, $\rho_0=2650$kg/m$^3$, and $T=1.5 \times 10^4$K. 
After this procedure, $\Delta {\rm PE}$ in the two EOS become similar.
Thus, although this approximation is simple, it provides a reasonable answer.

\subsection{The Rankine-Hugoniot equations}
\label{S.sugita}
\cite{Sugitaetal2012} derive the following differential equations to describe after-shock temperature $T$ and entropy $S$ based on the Rankine-Hugoniot equations;

\begin{equation}
\frac{dT}{dU_p}=C_0 \gamma_0 T \frac{(U_s-U_p)^{q-1}}{U_s^{q+1}}+\frac{s U_p^2}{C_V U_s},
\label{sugita1}
\end{equation} 
\begin{equation}
\frac{dS}{dU_p}=\frac{sU_p^2}{TU_s}.
\label{sugita1}
\end{equation} 

Here, $p=p_i+\rho_i U_s U_p$ and $\rho=\rho_i U_s/(U_s - U_p)$, where $p_i$, $\rho_i$, $U_s$, and $U_p$ are the pre-shock pressure, pre-shock density, shock velocity and particle velocity. $U_s$ and $U_p$ has a relation $U_s=C_0+s U_p$, where $C_0$ and $s$ are the sound speed and constant.
For our calculations, we choose $s=1.56$ (for MgSiO$_3$ bridgmanite, \citealt{Dengetal2008}) and $T_i=2000$K (pre-shock temperature). 
At $p_i=0$ GPa, $\rho_i=4100$kg/m$^3$, $C_0=6.47$km/s and at  $p_i=50$ GPa, $\rho_i=4500$kg/m$^3$, $C_0=9.0$km/s.

\subsection{Pressure vs. entropy increase}
\label{S.PressEn}
Figure \ref{fig:time_p_S} shows the relationship between the pressure (shown in grey) and entropy gain (shown in green).
We choose a specific SPH particle from each simulation and follow its properties.
In (a), the primary impact ($\sim 90$ GPa) is the major source for the entropy increase.  The entropy changes overtime, but the extent is limited.
In (b), the SPH particle is heated by multiple shocks, including the primary impact-induced shock and shocks due to the planetary expansions and contractions (discussed in Section 4.1).
After $\sim$ 5 hrs, the entropy slowly increases due to continuous small-scale planetary deformation (the planet continues to wobble) until the system reaches its equilibrium state. 
In (c), the SPH particle experiences a number of shocks because the target and impact collide several times. The entropy gain is larger than the other two cases.

\begin{figure*}
\setcounter{figure}{1}
\renewcommand{\thefigure}{S\arabic{figure}}

  \begin{center}
     \includegraphics[scale=1]{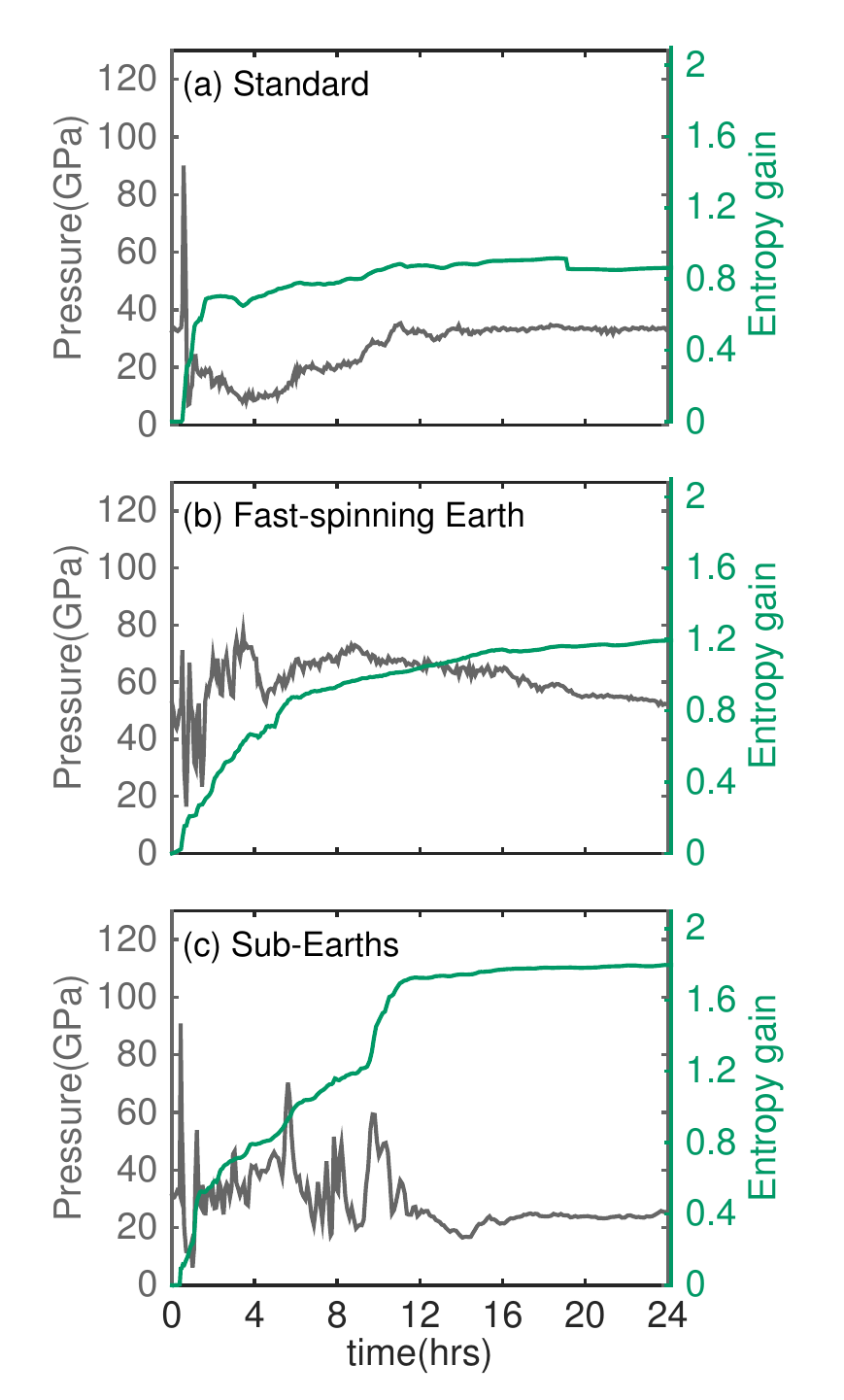}
  \end{center}
  \caption{Time dependence of the pressure and entropy (in 10$^3$ J/K/kg) of an SPH particle during the initial 24 hours.}
\label{fig:time_p_S}
\end{figure*}

\subsection{Further discussions on the mixing analysis}
\label{S.add.Mx}
We assume that the Earth's mantle was chemically heterogeneous before the impact, but here we further discuss its plausibility.
Unlike (a) or (c), the model (b) requires that the Earth spins very quickly before the giant impact.
This may indicate that the Earth experienced another (older) giant impact before the Moon-forming impact.
This is because the angular momentum of a planet delivered by a number of small impacts from random directions tend to cancel out.
This older giant impact could have been similar to the ``sub-Earths'' model, meaning that two similar mass objects collided, because this type of an impact is one of the easiest ways to deliver a large angular momentum to the planet \citep{Canup2014}. 
If this is the case, Earth's mantle could have been homogenized before the Moon-forming impact.
If the heterogeneity formation predated this older impact, this could be a potential problem for (b).
Alternatively, it is also possible that the heterogeneity formed between this older and the last giant impacts, possibly in the form of a basal magma ocean by fractional melting and crystallization processes. 
The re-establishment of a compositionally distinct basal magma ocean could have been accomplished in less than $10^6$ yr compared to the likely time between giant impacts plausibly $\sim 10^7$ years (without an atmosphere, the majority of the mantle could have crystallized as short as $10^3$ years, \citealt{Solomatov2000}).
It should be noted that there is no reason to suppose that this older impact was immediately prior, but the time interval might have been preferably short if the older impact formed a satellite. This is because the interaction between the Earth and satellite may have slowed the Earth's spin rate within $10^6-10^7$ years \citep{Canup2014}.
This older satellite might have merged with a newer satellite formed by the last giant impact \citep{Citronetal2014}.

Another potential problem is that the critical value $0.5$ has been derived to analyze the stability of horizontally stratified layers, but the value can differ for spherically stratified layers, as in our model. However, there is no literature concerning this geometry; thus, we simply apply the critical value for our analyses. The choice of the minimum density could also affect the estimate of $\Delta {\rm PE}$.

Furthermore, we only perform one simulation for each model and EOS. It is possible that $\Delta {\rm KE}$ and $\Delta {\rm PE}$ can change even for the same type of impact depending on the choice of the initial conditions (e.g., $v_{\rm imp}$ and $b$). 
To perform a simple and crude analysis, here we assume that the planetary kinetic energy is expressed as $\frac{1}{2}I \omega^2$, where $I$ is the moment of inertia and $\omega$ is the angular velocity of the planet, and that $I$ and $\Delta {\rm PE}$ do not vary in the same model. 
We compute the ratio of $\Delta {\rm PE}/\Delta {\rm KE}$ based on $\omega$ from published successful simulations \citep{CukStewart2012, Canup2012} and find that most of these simulations do not change the ratio large enough to cross the critical value 0.5, except Run 14 ($M_{\rm i}/M_{\rm T}$=0.45, $b$=0.40, and $v_{\rm imp}/v_{\rm esc}=1.4$) in the sub-Earths model, depending on the EOS (0.52 for MgSiO$_3$ liquid and 0.38 for forsterite).
Thus, our results likely provide the general trend, but some outlier may exist.
Nevertheless, the choice of initial conditions is not likely to alter the signatures of $dS/dr>0$ or the melting of the nearly entire mantle because these are states less sensitive to the conditions.



\bibliographystyle{elsarticle-harv}
\bibliography{miki_mantle.bib2.bib}

\begin{thebibliography}{60}
\expandafter\ifx\csname natexlab\endcsname\relax\def\natexlab#1{#1}\fi
\expandafter\ifx\csname url\endcsname\relax
  \def\url#1{\texttt{#1}}\fi
\expandafter\ifx\csname urlprefix\endcsname\relax\def\urlprefix{URL }\fi

\bibitem[{Abe and Matsui(1986)}]{AbeMatsui1986}
Abe, Y., Matsui, T., 1986. {Early evolution of the Earth: Accretion, atmosphere
  formation, and thermal history}. J. Geophys. Res. 91, E291.

\bibitem[{Armytage et~al.(2012)Armytage, Georg, Williams, and
  Halliday}]{Armytageetal2012}
Armytage, R. M.~G., Georg, R.~B., Williams, H.~M., Halliday, A.~N., 2012.
  {Silicon isotopes in lunar rocks: Implications for the Moon's formation and
  the early history of the Earth}. Geochim. Cosmochia. Ac. 77, 504--514.

\bibitem[{Benz et~al.(1986)Benz, Slattery, and Cameron}]{Benzetal1986}
Benz, W., Slattery, W.~L., Cameron, A. G.~W., 1986. {The origin of the Moon and
  the single-impact hypothesis. I}. Icarus 66, 515--535.

\bibitem[{Cameron and Ward(1976)}]{CameronWard1976}
Cameron, A. G.~W., Ward, W.~R., 1976. {The origin of the Moon}. Lunar Planet.
  Sci. VII, 120.

\bibitem[{Canup(2008)}]{Canup2008b}
Canup, R.~M., 2008. {Accretion of the Earth}. Philos Trans A Math Phys Eng Sci
  366, 4061--75.

\bibitem[{Canup(2012)}]{Canup2012}
Canup, R.~M., 2012. {Forming a Moon with an Earth-like composition via a giant
  impact}. Science 338, 1052--1055.

\bibitem[{Canup(2014)}]{Canup2014}
Canup, R.~M., 2014. {Lunar-forming impacts: processes and alternatives}. Phil.
  Trans. R. Soc. A 372~(2024), 20130175.

\bibitem[{Carlson and Lugmair(1988)}]{CarlsonLugmair1988}
Carlson, R.~W., Lugmair, G.~W., 1988. {The age of ferroan anorthosite 60025 -
  Oldest crust on a young moon?} Earth Planet. Sci. Lett. 90, 119--130.

\bibitem[{Chandrasekhar(1961)}]{Chandrasekhar1961}
Chandrasekhar, S., 1961. {Hydrodynamic and hydromagnetic stability}. Oxford
  Univ. Press, Oxford.

\bibitem[{Citron et~al.(2014)Citron, Aharonson, Perets, and
  Genda}]{Citronetal2014}
Citron, R.~I., Aharonson, O., Perets, H., Genda, H., 2014. {Moon formation from
  multiple large impacts.} Lunar Planet. Sci 45th, 2085.

\bibitem[{\'{C}uk and Stewart(2012)}]{CukStewart2012}
\'{C}uk, M., Stewart, S.~T., 2012. {Making the Moon from a fast-spinning Earth:
  A giant impact followed by resonant despinning}. Science 338, 1047--1052.

\bibitem[{Dahl and Stevenson(2010)}]{DahlStevenson2010}
Dahl, T.~W., Stevenson, D.~J., 2010. {Turbulent mixing of metal and silicate
  during planet accretion - and interpretation of the Hf–W chronometer}.
  Earth Planet. Sci. Lett. 295, 177--186.

\bibitem[{Dauphas et~al.(2014)Dauphas, Burkhardt, Warren, and
  Fang-Zhen}]{Dauphasetal2014}
Dauphas, N., Burkhardt, C., Warren, P.~H., Fang-Zhen, T., 2014. {Geochemical
  arguments for an Earth-like Moon-forming impactor}. Phil. Trans. R. Soc. A
  372, 20130244.

\bibitem[{de~Koker and Stixrude(2009)}]{deKokerStixrude2009}
de~Koker, N., Stixrude, L., 2009. {Self-consistent thermodynamic description of
  silicate liquids, with application to shock melting of MgO periclase and
  MgSiO3 perovskite}. Geophysical J. Int. 178, 162--179.

\bibitem[{de~Vries et~al.(2014)de~Vries, Nimmo, Jacobson, Morbidelli, and
  Rubie}]{deVriesetal2014}
de~Vries, J., Nimmo, F., Jacobson, S.~A., Morbidelli, A., Rubie, D.~C., 2014.
  {Melting Due to Impacts on Growing Proto-Planets}. Lunar Planet. Sci. 45th,
  1896.

\bibitem[{Deng et~al.(2008)Deng, Gong, and Fei}]{Dengetal2008}
Deng, L., Gong, Z., Fei, Y., 2008. {Direct shock wave loading of MgSiO3
  perovskite to lower mantle conditions and its equation of state}. Phys. Earth
  Planet. In. 170, 210--214.

\bibitem[{Elkins-Tanton(2008)}]{ElkinsTanton2008}
Elkins-Tanton, L.~T., 2008. {Linked magma ocean solidification and atmospheric
  growth for Earth and Mars}. Earth Planet. Sci. Lett. 271, 181--191.

\bibitem[{Fiquet et~al.(2000)Fiquet, Dewaele, and Andrault}]{Fiquetetal2000}
Fiquet, G., Dewaele, A., Andrault, D., 2000. {Thermoelastic properties and
  crystal structure of MgSiO$_3$ perovskite at lower mantle pressure and
  temperature conditions}. Geophys. Res. Lett. 27, 21--24.

\bibitem[{Fitoussi and Bourdon(2012)}]{FitoussiBourdon2012}
Fitoussi, C., Bourdon, B., 2012. {Silicon Isotope Evidence Against an Enstatite
  Chondrite Earth}. Science 335, 1477--1480.

\bibitem[{Franchi et~al.(1999)Franchi, Wright, Sexton, and
  Pillinger}]{Franchietal1999}
Franchi, I.~A., Wright, I.~P., Sexton, A.~S., Pillinger, C.~T., 1999. The
  oxygen-isotopic composition of earth and mars. Meteoritics and Planetary
  Science 34, 657--661.

\bibitem[{Hartmann and Davis(1975)}]{HartmannDavis1975}
Hartmann, W.~K., Davis, D.~R., 1975. {Satellite-sized planetesimals and lunar
  origin}. Icarus 24, 504--514.

\bibitem[{Herwartz et~al.(2014)Herwartz, Pack, Friedrichs, and
  Bischoff}]{Herwartzetal2014}
Herwartz, D., Pack, A., Friedrichs, B., Bischoff, A., 2014. {Identification of
  the giant impactor Theia in lunar rocks}. Science 344, 1146--50.

\bibitem[{Jacobson et~al.(2014)Jacobson, Morbidelli, Raymond, O'Brien, Walsh,
  and Rubie}]{Jacobsonetal2014}
Jacobson, S.~A., Morbidelli, A., Raymond, S.~N., O'Brien, D.~P., Walsh, K.~J.,
  Rubie, D.~C., 2014. {Highly siderophile elements in Earth's mantle as a clock
  for the Moon-forming impact}. Nature 508, 84--87.

\bibitem[{Kaib and Cowan(2015)}]{KaibCowan2015}
Kaib, N.~A., Cowan, N.~B., 2015. {The feeding zones of terrestrial planets and
  insights into Moon formation}. Icarus 252, 161--174.

\bibitem[{Karato(2014)}]{Karato2014}
Karato, S., 2014. {Asymmetric shock heating and the terrestrial magma ocean
  origin of the Moon}. Proceedings of the Japan Academy, Series B 90, 97--103.

\bibitem[{Labrosse et~al.(2007)Labrosse, Hernlund, and
  Coltice}]{Labrosseetal2007}
Labrosse, S., Hernlund, J.~W., Coltice, N., 2007. {A crystallizing dense magma
  ocean at the base of the Earth's mantle}. Nature 450, 866--869.

\bibitem[{Lee et~al.(1997)Lee, Halliday, Snyder, and Taylor}]{Leeetal1997}
Lee, D.-C., Halliday, A., Snyder, G.~A., Taylor, L.~A., 1997. {Age and Origin
  of the Moon}. Science 278, 1098--1103.

\bibitem[{Lee and Halliday(1996)}]{LeeHalliday1996}
Lee, D.-C., Halliday, A.~N., 1996. {Hf-W Isotopic Evidence for Rapid Accretion
  and Differentiation in the Early Solar System}. Science 274, 1876--1879.

\bibitem[{Li and Agee(1996)}]{LiAgee1996}
Li, J., Agee, C.~B., 1996. {Geochemistry of mantle-core differentiation at high
  pressure}. Nature 381, 686--689.

\bibitem[{Mastrobuono-Battisti et~al.(2015)Mastrobuono-Battisti, Perets, and
  Raymond}]{MastrobuonoBattistietal2015}
Mastrobuono-Battisti, A., Perets, H.~B., Raymond, S.~N., 2015. {A primordial
  origin for the compositional similarity between the Earth and the Moon}.
  Nature 520, 212--215.

\bibitem[{Melosh(2007)}]{Melosh2007}
Melosh, H.~J., 2007. {A hydrocode equation of state for SiO2}. Meteorit.
  Planet. Sci. 42, 2079--2098.

\bibitem[{Mukhopadhyay(2012)}]{Mukhopadhyay2012}
Mukhopadhyay, S., 2012. {Early differentiation and volatile accretion recorded
  in deep-mantle neon and xenon}. Nature 486, 101--104.

\bibitem[{Nakajima and Stevenson(2014)}]{NakajimaStevenson2014}
Nakajima, M., Stevenson, D.~J., 2014. {Investigation of the Initial State of
  the Moon-Forming Disk: Bridging SPH Simulations and Hydrostatic Models}.
  Icarus 233, 259--267.

\bibitem[{Pahlevan and Stevenson(2007)}]{PahlevanStevenson2007}
Pahlevan, K., Stevenson, D.~J., 2007. {Equilibration in the aftermath of the
  lunar-forming giant impact}. Earth Planet. Sci. Lett. 262, 438--449.

\bibitem[{Pahlevan et~al.(2011)Pahlevan, Stevenson, and
  Eiler}]{Pahlevanetal2011}
Pahlevan, K., Stevenson, D.~J., Eiler, J.~M., 2011. {Chemical fractionation in
  the silicate vapor atmosphere of the Earth}. Earth Planet. Sci. Lett. 301,
  433--443.

\bibitem[{Reufer et~al.(2012)Reufer, Meier, Benz, and Wieler}]{Reuferetal2012}
Reufer, A., Meier, M. M.~M., Benz, W., Wieler, R., 2012. {A hit-and-run giant
  impact scenario}. Icarus 221, 296--299.

\bibitem[{Righter et~al.(1997)Righter, Drake, and Yaxley}]{Righteretal1997}
Righter, K., Drake, M.~J., Yaxley, G., 1997. {Prediction of siderophile element
  metal-silicate partition coefficients to 20 GPa and 2800°C: the effects of
  pressure, temperature, oxygen fugacity, and silicate and metallic melt
  compositions}. Phys. Earth Planet. Inter. 100~(1), 115--134.

\bibitem[{Rubie et~al.(2015)Rubie, Jacobson, Morbidelli, O'Brien, Young,
  de~Vries, Nimmo, Palme, and Frost}]{Rubieetal2015}
Rubie, D.~C., Jacobson, S.~A., Morbidelli, A., O'Brien, D.~P., Young, E.~D.,
  de~Vries, J., Nimmo, F., Palme, H., Frost, D.~J., 2015. {Accretion and
  differentiation of the terrestrial planets with implications for the
  compositions of early-formed Solar System bodies and accretion of water}.
  Icarus 248, 89--108.

\bibitem[{Rubie et~al.(2003)Rubie, Melosh, Reid, Liebske, and
  Righter}]{Rubieetal2003}
Rubie, D.~C., Melosh, H.~J., Reid, J.~E., Liebske, C., Righter, K., 2003.
  {Mechanisms of metal-silicate equilibration in the terrestrial magma ocean}.
  Earth Planet. Sci. Lett. 205, 239--255.

\bibitem[{Shi et~al.(2013)Shi, Zhang, Yang, Liu, Wang, Meng, Andrews, and
  Mao}]{Shietal2013}
Shi, C.~Y., Zhang, L., Yang, W., Liu, Y., Wang, J., Meng, Y., Andrews, J.~C.,
  Mao, W.~L., 2013. {Formation of an interconnected network of iron melt at
  Earth's lower mantle conditions}. Nat. Geosci. 6, 971--975.

\bibitem[{Solomatov(2000)}]{Solomatov2000}
Solomatov, V.~S., 2000. {Fluid Dynamics of a Terrestrial Magma Ocean}. Origin
  of the Earth and Moon. Univ.of Arizona Press, Tucson, pp. 323--338.

\bibitem[{Stevenson(1989)}]{Stevenson1989}
Stevenson, D.~J., 1989. {Spontaneous small-scale melt segregation in partial
  melts undergoing deformation}. Geophys. Res. Lett. 16, 1067--1070.

\bibitem[{Stevenson(1990)}]{Stevenson1990}
Stevenson, D.~J., 1990. {Fluid dynamics of core formation}. Newton, H. E. and
  Jones, J.H. (Ed.), Origin of the Earth. Oxford University Press, New York,
  pp. 231 - 250.

\bibitem[{Stewart et~al.(2014)Stewart, Lock, and
  Mukhopadhyay}]{Stewartetal2014}
Stewart, S.~T., Lock, S., Mukhopadhyay, S., 2014. {Partial atmospheric loss and
  partial mantle melting during the giant impact stage of planet formation}.
  AGU Meeting, San Francisco, P44A--06.

\bibitem[{Stixrude et~al.(2009)Stixrude, de~Koker, Sun, Mookherjee, and
  Karki}]{Stixrudeetal2009}
Stixrude, L., de~Koker, N., Sun, N., Mookherjee, M., Karki, B., B., 2009.
  {Thermodynamics of silicate liquids in the deep Earth}. Earth Planet. Sci.
  Lett. 278, 226--232.

\bibitem[{Stixrude and Karki(2005)}]{StixrudeKarki2005}
Stixrude, L., Karki, B., 2005. {Structure and Freezing of MgSiO$_3$ Liquid in
  Earth's Lower Mantle}. Science 310, 297--299.

\bibitem[{Sugita et~al.(2012)Sugita, Kurosawa, Kadono, and
  Sano}]{Sugitaetal2012}
Sugita, S., Kurosawa, K., Kadono, T., Sano, T., 2012. {An High-Precistion
  Semi-Analytical on-Hugoniot EOS for Geologic Materials}. Lunar Planet. Sci.
  43rd, 2053.

\bibitem[{Tackley(2012)}]{Tackley2012}
Tackley, P.~J., 2012. {Dynamics and evolution of the deep mantle resulting from
  thermal, chemical, phase and melting effects}. Earth-Science Reviews 110,
  1--25.

\bibitem[{Taylor(1931)}]{Taylor1931}
Taylor, G.~I., 1931. {Effect of Variation in Density on the Stability of
  Superposed Streams of Fluid}. Proc. R. Soc. London A 132, 499--523.

\bibitem[{Thompson and Lauson(1972)}]{ThompsonLauson1972}
Thompson, S.~L., Lauson, H.~S., 1972. {Improvements in the CIIARTD radiation
  hydrodynamics code III: revised analytic equation of state}. Sandia National
  Laboratories, Albuquerque, New Mexico, 119p.

\bibitem[{Tonks and Melosh(1993)}]{TonksMelosh1993}
Tonks, W.~B., Melosh, J., 1993. {Magma Ocean Formation Due to Giant Impact}. J.
  Geophys. Res. 98, 5319--5333.

\bibitem[{Touboul et~al.(2007)Touboul, Kleine, B., Palme, and
  Wieler}]{Toubouletal2007}
Touboul, M., Kleine, T., B., B., Palme, H., Wieler, R., 2007. {Late formation
  and prolonged differentiation of the Moon inferred from W isotopes in lunar
  metals}. Nature 450, 1206--1209.

\bibitem[{Touboul et~al.(2012)Touboul, Puchtel, and Walker}]{Toubouletal2012}
Touboul, M., Puchtel, I.~S., Walker, R.~J., 2012. {182W Ecidence for Long-Term
  Preservation of Early Mantle Differentiation Products}. Science 335,
  1065--1069.

\bibitem[{Touma and Wisdom(1998)}]{ToumaWisdom1998}
Touma, J., Wisdom, J., 1998. {Resonances in the Early evolution of the
  Earth-Moon system}. Astron. J. 115, 1653--1663.

\bibitem[{Wade and Wood(2005)}]{WadeWood2005}
Wade, J., Wood, B.~J., 2005. {Core formation and the oxidation state of the
  Earth}. Earth Planet. Sci. Lett. 236, 78--95.

\bibitem[{Walsh et~al.(2011)Walsh, Morbidelli, Raymond, O'Brien, and
  Mandell}]{Walshetal2011}
Walsh, K.~J., Morbidelli, A., Raymond, S.~N., O'Brien, D.~P., Mandell, A.~M.,
  2011. {The Low Mass of Mars: First Evidence of Early Gas-Driven Migration}.
  Nature 475, 206--209.

\bibitem[{Wiechert et~al.(2001)Wiechert, Halliday, Lee, Snyder, Taylor, and
  Rumble}]{Wiecheretal2001}
Wiechert, U., Halliday, A.~N., Lee, D.~C., Snyder, G.~A., Taylor, L.~A.,
  Rumble, D., 2001. {Oxygen isotopes and the Moon-forming giant impact}.
  Science 294, 345--348.

\bibitem[{Willbold et~al.(2011)Willbold, Elliott, and
  Moorbath}]{Willboldetal2011}
Willbold, M., Elliott, T., Moorbath, S., 2011. {The tungsten isotopic
  composition of the Earth's mantle before the terminal bombardment}. Nature
  477, 195--199.

\bibitem[{Wisdom and Tian(2015)}]{WisdomTian2015}
Wisdom, J., Tian, Z., 2015. {Early evolution of the EarthÐMoon system with a
  fast-spinning Earth}. Icarus 256, 138--146.

\bibitem[{Zimmerman et~al.(1999)Zimmerman, Zhang, Kohlstedt, and
  Karato}]{Zimmermanetal1999}
Zimmerman, M.~E., Zhang, S., Kohlstedt, D.~L., Karato, S., 1999. {Melt
  distribution in mantle rocks deformed in shear}. Geophys. Res. Lett. 26,
  1505--1508.

\end{thebibliography}

\end{document}